\documentclass[12pt]{article}
\usepackage{graphicx}
\usepackage{amsmath}
\usepackage{amssymb}
\usepackage{caption2}
\begin{document}
\bibliographystyle{prsty}
\begin{center}
{\large {\bf \sc{  Analysis of the vertices $D^*D_sK$, $D^*_sDK$,
$D_0D_sK$ and $D_{s0}DK$ with   the light-cone QCD sum rules  }}} \\[2mm]
Z. G. Wang$^{1}$ \footnote{Corresponding author; E-mail,wangzgyiti@yahoo.com.cn.  }, S. L. Wan$^{2}$     \\
$^{1}$ Department of Physics, North China Electric Power University, Baoding 071003, P. R. China \\
$^{2}$ Department of Modern Physics, University of Science and Technology of China, Hefei 230026, P. R. China \\
\end{center}

\begin{abstract}
In this article, we analyze the vertices $D^*D_sK$, $D^*_sDK$,
$D_0D_sK$ and $D_{s0}DK$ within the framework of the light-cone QCD
sum rules approach in an unified way. The strong coupling constants
$G_{D^*D_sK}$ and $G_{D^*_sDK}$ are   important  parameters in
evaluating the charmonium absorption cross sections in searching for
the quark-gluon plasmas, our numerical values of the $G_{D^*D_sK}$
and $G_{D^*_sDK}$ are compatible with the existing estimations
although somewhat smaller, the $SU(4)$ symmetry breaking effects are
very large, about $60\%$. For the charmed scalar mesons $D_0$ and
$D_{s0}$,  we  take the point of view that  they are the
conventional $c\bar{u}$ and $c\bar{s}$ mesons  respectively, and
calculate the strong coupling constants $G_{D_0 D_s K}$ and
$G_{D_{s0} D K}$ with the vector interpolating currents. The
numerical values of the scalar-$D_sK$ and -$DK$ coupling constants
$G_{D_0 D_s K}$ and $G_{D_{s0} D K}$ are compatible with the
existing estimations, the large values support the hadronic dressing
mechanism.  Furthermore, we study the dependence  of the four strong
coupling constants $G_{D^*D_sK}$, $G_{D^*_sDK}$, $G_{D_0D_sK}$ and
$G_{D_{s0}DK}$  on the non-perturbative parameter $a_4$ of the
twist-2 $K$ meson light-cone distribution amplitude.
\end{abstract}


\section{Introduction}

The suppression of  the $J/\psi$ production in  relativistic  heavy
ion collisions maybe  one of the important signatures to identify
the possible phase  transition  to the quark-gluon plasma
\cite{Matsui86}. The dissociation of the $J/\psi$ in the quark-gluon
plasma due to color screening can lead to a reduction of its
production, however, the $J/\psi$ suppression maybe already present
in the hadron-nucleus collisions. It is necessary to separate  the
absorption of the $J/\psi$ by the nucleons and by the co-mover light
mesons ($\pi$, $K$, $\rho$, $\omega$, etc.)   before we can make a
definitive conclusion about the formation of the quark-gluon plasma.
It is of great importance to understand the $J/\psi$ production and
absorption mechanisms in the hadronic matter.   The values of the
$J/\psi$ absorption cross sections by the light hadrons  are not
known empirically, we have to resort to some theoretical approaches.
Among existing  approaches for evaluating the charmonium absorption
cross sections by the light hadrons,  the one-meson exchange model
and the effective $SU(4)$ theory are typical \cite{MesonEx, SU4}.
The detailed knowledge  about the hadronic vertices  or the strong
coupling constants which are basic parameters in the effective
Lagrangians is of great importance.

The  discovery of the two strange-charmed mesons $D_{s0} $ and
$D_{s1}  $ with spin-parity $0^+$ and $1^+$ respectively has
triggered hot  debate on their nature, under-structures and whether
it is necessary to introduce  the exotic  states \cite{exp03}. The
mass of the $D_{s0} $ is significantly lower than the values of the
$0^+$ state mass  from the quark models   and lattice simulations
\cite{QuarkLattice}. The difficulties to identify the $D_{s0} $ and
$D_{s1} $ states with the conventional $c\overline{s}$ mesons are
rather similar to those appearing in the light scalar mesons below
$1GeV$. Among the various explanations,
 the hadronic dressing mechanism is typical.  The scalar
mesons $a_0(980)$, $f_0(980)$, $D_0$ and $D_{s0}$ may have bare
$q\overline{q}$, $\bar{c}u$ and  $c\bar{s}$ kernels in the $P-$wave
states with strong coupling  to the nearby threshold respectively,
the $S-$wave virtual intermediate hadronic states (or the virtual
mesons loops) play a crucial role in the composition of those bound
states (or resonances due to the masses below or above the
thresholds). The hadronic dressing mechanism (or unitarized quark
models) takes the point of view that the $f_0(980)$, $a_0(980)$ ,
$D_0$ and $D_{s0}$ mesons have small $q\bar{q}$, $\bar{c}u$ and
$c\bar{s}$  kernels of the typical $q\bar{q}$, $c\bar{u}$ and
$c\bar{s}$ mesons size respectively. The strong couplings to the
virtual intermediate hadronic states (or the virtual mesons loops)
may result in smaller masses than the conventional scalar $q\bar{q}$
, $c\bar{u}$ and $c\bar{s}$ mesons in the constituent quark models,
enrich the pure $q\bar{q}$ , $c\bar{u}$ and $c\bar{s}$ states with
other components \cite{HDress,UQM}. Those mesons may spend part (or
most part) of their lifetime as virtual $ K \bar{K} $, $D_sK$ and
$DK$ states \cite{HDress,UQM}.   It is interesting to study the
possibility of the hadronic dressing mechanism.

In this article, we calculate the values of the strong coupling
constants $G_{D^* D_s K}$, $G_{D_s ^* D K}$, $G_{D_0 D_s K}$ and
$G_{D_{s0} D K}$ within the framework of the light-cone QCD sum
rules approach. The light-cone QCD sum rules approach carries out
the operator product expansion near the light-cone $x^2\approx 0$
instead of the short distance $x\approx 0$ while the
non-perturbative matrix elements  are parameterized by the
light-cone distribution amplitudes
 which classified according to their twists  instead of
 the vacuum condensates \cite{LCSR,LCSRreview}.
Furthermore, we study the dependence of the strong coupling
constants $G_{D^* D_s K}$, $G_{D_s ^* D K}$, $G_{D_0 D_s K}$ and
$G_{D_{s0} D K}$ on the coefficient $a_4$ of the twist-2 $K$ meson
light-cone distribution amplitude $\phi_K(u)$, and estimate the
values of the non-perturbative parameter. It is very difficult to
determine the $a_4$ with the QCD sum rules, the values of the $a_4$
suffer from large uncertainties, as   it concerns  high dimension
vacuum condensates which are known poorly
\cite{LCSR,LCSRreview,Belyaev94,Ball98,Ball06}. It is of great
importance to determine the values directly from the  experimental
data.

The article is arranged as: in Section 2, we derive the strong
coupling constants  $G_{D^* D_s K}$, $G_{D^*_s D K}$, $G_{D_0 D_s
K}$ and $G_{D_{s0} D K}$ within the framework of the light-cone QCD
sum rules approach; in Section 3, the numerical results and
discussions; and in Section 4, conclusion.

\section{Strong coupling constants  $G_{D^* D_s K}$, $G_{D^*_s D K}$, $G_{D_0 D_s K}$
and $G_{D_{s0} D K}$   with light-cone QCD sum rules}

In the following, we write down the definitions  for the strong
coupling constants $G_{D^* D_s K}$, $G_{D^*_s D K}$, $G_{D_0 D_s K}$
and $G_{D_{s0} D K}$ ,
\begin{eqnarray}
\langle D^*(q+P) D_s(q)| K(P)\rangle&=&G_{D^* D_s K} (P-q)\cdot
\epsilon \, ,
\nonumber\\
\langle D^*_s(q+P) D(q)| K(P)\rangle&=&G_{D^*_s D K} (P-q)\cdot
\epsilon \, ,
 \nonumber\\ \langle D_{0}(q+P) D_s(q)| K(P)\rangle&=&G_{D_0 D_s K} \,
 ,\nonumber\\
 \langle
D_{s0}(q+P) D(q)| K(P)\rangle&=&G_{D_{s0} D K} \, ,
\end{eqnarray}
here the $\epsilon_\mu$ are the polarization vectors  of the
 mesons $D^*$ and $D^*_s$.  We study the strong coupling constants $G_{D^* D_s
K}$, $G_{D^*_s D K}$, $G_{D_0 D_s K}$ and $G_{D_{s0} D K}$ with the
interpolating currents $ J_{D_{s}}(x)$, $J_{D}(x)$, $
J^{D_s}_\mu(x)$  and $J^D_\mu(x)$ in an unified way, and choose the
 two-point correlation functions $\Pi_{\mu}^1(P,q)$ and $\Pi_{\mu}^2(P,q)$,
\begin{eqnarray}
\Pi_{\mu}^1(P,q)&=&i \int d^4x \, e^{-i q \cdot x} \,
\langle 0 |T\left\{J^D_\mu(0) J_{D_{s}}(x)\right\}|K(P)\rangle \, , \\
\Pi_{\mu}^2(P,q)&=&i \int d^4x \, e^{-i q \cdot x} \,
\langle 0 |T\left\{J^{D_s}_\mu(0) J_{D}(x)\right\}|K(P)\rangle \, , \\
J^D_\mu(x)&=&{\bar u}(x)\gamma_\mu  c(x)\, ,\nonumber \\
J^{D_s}_\mu(x)&=&{\bar s}(x)\gamma_\mu  c(x)\, ,\nonumber\\
J_{D}(x)&=& \bar{c}(x)i\gamma_5u(x) \, , \nonumber\\
 J_{D_{s}}(x)&=& \bar{c}(x)i\gamma_5s(x) \, .
\end{eqnarray}
 The correlation functions
$\Pi^{1(2)}_{\mu}(P,q)$ can be decomposed as
\begin{eqnarray}
\Pi^{1(2)}_{\mu}(P,q)&=&\Pi^{1(2)}_{P}\left(q^2,(q+P)^2\right)P_{\mu}+\Pi^{1(2)}_{q}
\left(q^2,(q+P)^2\right)q_{\mu},
\end{eqnarray}
due to the Lorentz covariance. In this article, we derive the sum
rules with the tensor structures  $  P_\mu$  and $  q_\mu$
respectively, and make detailed studies.

According to the basic assumption of current-hadron duality in the
QCD sum rules approach \cite{SVZ79}, we can insert  a complete
series of intermediate states with the same quantum numbers as the
current operators $J_{D_{s}}(x)$ ($J_D(x)$) and $J^D_\mu(x)$
($J^{D_s}_\mu(x)$) into the correlation function $\Pi^1_{\mu} $
($\Pi^2_{\mu} $) to obtain the hadronic representation. After
isolating the ground states and the first orbital excited states
contributions from the pole terms of the $D_{s}$, $D^*$ and $D_0$
($D$, $D^*_s$ and $D_{s0}$ )  mesons, the correlation function $
\Pi^1_\mu$ ($\Pi^2_\mu$) can be expressed in terms of the strong
coupling constants $G$ and the  decay constants $f_M$ of the heavy
mesons, the explicit expressions are presented in the appendix. We
use the standard definitions  for the   decay constants $f_M$
($f_{D_s}$, $f_D$, $f_{D_s^*}$, $f_{D^*}$, $f_{D_{s0}}$, $f_{D_{0}}$
) of the heavy mesons,
\begin{eqnarray}
\langle0|J_{D}(0)|D(q)\rangle&=& \frac{f_{D}m_D^2}{m_c+m_u} \, ,\nonumber \\
\langle0|J_{D_{s}}(0)|D_{s}(q)\rangle&=& \frac{f_{D_s}m_{D_s}^2}{m_c+m_s}\, ,\nonumber \\
\langle0|J^{D}_\mu(0)|D^*(q)\rangle&=& f_{D^*}m_{D^*}\epsilon_\mu \, ,\nonumber\\
\langle0|J^{D_s}_\mu(0)|D^*_{s}(q)\rangle&=& f_{D_s^*}m_{D_s^*}\epsilon_\mu \, ,\nonumber\\
\langle0|J^{D}_\mu(0)|D_0(q)\rangle&=& f_{D_0}q_\mu \, , \nonumber\\
\langle0|J^{D_s}_\mu(0)|D_{s0}(q)\rangle&=& f_{D_{s0}}q_\mu \, .
\end{eqnarray}

The  quarks $c$ and $s$ have finite and non-equal masses, the
non-conservation  of the vector currents $J^{D_s}_\mu(x)$ and
$J^D_\mu(x)$ can lead to the non-vanishing couplings  to the scalar
mesons $D_{s0}$ and $D_0$ beside the vector mesons $D^*_s$ and
$D^*$, we can study the properties of those mesons with the two
interpolating currents $J^{D_s}_\mu(x)$ and $J^D_\mu(x)$ in an
unified way.
 Here we have not shown the contributions from
the high resonances and  continuum states explicitly as they are
suppressed due to the double Borel transformation. The numerical
values of the fractions
\begin{eqnarray}
\frac{m_{D^*}^2-m_{D_s}^2+m_K^2}{m_{D^*}^2} \, , \,
\frac{m_{D^*_s}^2-m_D^2+m_K^2}{m_{D^*_s}^2}  \nonumber
\end{eqnarray}
are less than $30\%$ and the corresponding spectral densities for
the ground states are greatly suppressed, the tensor structures with
$q_{\mu}$ are especially suitable for studying the first orbital
excited states $D_0$ and $D_{s0}$ with the vector currents. The
numerical values of the fractions
\begin{eqnarray}
\frac{m_{D^*}^2+m_{D_s}^2-m_K^2}{m_{D^*}^2} \, , \,
\frac{m_{D^*_s}^2+m_D^2-m_K^2}{m_{D^*_s}^2} \nonumber
\end{eqnarray}
are about $2$, the  tensor structures with $P_{\mu}$ are especially
suitable for studying the ground states $D^*$ and $D^*_s$ with the
vector currents.

Now we carry out the operator product expansion near the light-cone
$x^2\approx 0$ to obtain the    representation at the level of
quark-gluon degrees of freedom for the correlation functions
$\Pi^1_\mu$ and $\Pi^2_\mu$. In the following, we briefly outline
the operator product expansion for the correlation functions
$\Pi^1_\mu$ and $\Pi^2_\mu$ in perturbative QCD theory. The
calculations are performed at the large space-like momentum regions
$(q+P)^2\ll 0$ and  $q^2\ll 0$, which correspond to the small
light-cone distance $x^2\approx 0$ required by the validity of the
operator product expansion approach. We write down the propagator of
a massive quark in the external gluon field in the Fock-Schwinger
gauge firstly \cite{Belyaev94},
\begin{eqnarray}
&&\langle 0 | T \{q_i(x_1)\, \bar{q}_j(x_2)\}| 0 \rangle =
 i \int\frac{d^4k}{(2\pi)^4}e^{-ik(x_1-x_2)}\nonumber\\
 &&\left\{
\frac{\not\!k +m}{k^2-m^2} \delta_{ij} -\int\limits_0^1 dv\, g_s \,
G^{\mu\nu}_{ij}(vx_1+(1-v)x_2)
 \right. \nonumber \\
&&\left. \Big[ \frac12 \frac {\not\!k
+m}{(k^2-m^2)^2}\sigma_{\mu\nu} - \frac1{k^2-m^2}v(x_1-x_2)_\mu
\gamma_\nu \Big]\right\}\, ,
\end{eqnarray}
here $G^{\mu \nu }$ is the gluonic field strength, $g_s$ denotes the
strong coupling constant. Substituting the above $c$  quark
propagator and the corresponding $K$ meson light-cone distribution
amplitudes into the correlation functions $\Pi^1_\mu$ and
$\Pi^2_\mu$ in Eqs.(2-3) and completing the integrals over the
variables $x$ and $k$, finally we obtain the representation at the
level of quark-gluon degrees of freedom, the explicit expressions
are presented in the appendix. In calculation, we have used the
two-particle and three-particle $K$ meson light-cone distribution
amplitudes \cite{LCSR,LCSRreview,Belyaev94,Ball98,Ball06}, the
explicit expressions are also presented  in the appendix. The
parameters in the light-cone distribution amplitudes are scale
dependent and can be estimated with the QCD sum rules approach
\cite{LCSR,LCSRreview,Belyaev94,Ball98,Ball06}. In this article, the
energy scale $\mu$ is chosen to be  $\mu=1GeV$.

We perform the double Borel transformation with respect to  the
variables $Q_1^2=-(q+P)^2$  and $Q_2^2=-q^2$  for the correlation
functions $\Pi_P^{1(2)}$ and $\Pi_q^{1(2)}$, and obtain the
analytical expressions for those invariant functions, the explicit
expressions are presented in the appendix.

 In order to match the duality regions below the
thresholds $s_0$ and $s_0'$ for the interpolating currents
$J^{D}_{\mu}(x)$($J^{D_s}_{\mu}(x)$)    and $J_{D_s}(x)$ ($J_D(x)$ )
respectively, we can express the correlation functions
$\Pi_P^{1(2)}$ and $\Pi_q^{1(2)}$ at the level of quark-gluon
degrees of freedom into the following form,
\begin{eqnarray}
\Pi_{P(q)}^{1(2)}(q^2,(q+P)^2)&=& \int ds \int ds^\prime
\frac{\rho(s,s^\prime)}{\left\{ s-(q+P)^2\right \} (s^\prime-q^2) }
\, ,
\end{eqnarray}
then we  perform the double Borel transformation with respect to the
variables $Q_1^2=-(q+P)^2$  and $Q_2=-q^2$  directly. However, the
analytical expressions for the spectral densities $\rho(s,s')$ are
hard to obtain, we have to resort to  some approximations.  As the
contributions
 from the higher twist terms  are  suppressed by more powers of
 $\frac{1}{-q^2}$ or $\frac{1}{-(q+P)^2}$, the continuum subtractions will not affect the results remarkably,
here we will use the expressions in Eqs.(28-29) for the
three-particle (quark-antiquark-gluon) twist-3, twist-4  terms, and
the two-particle twist-4 terms. In fact, their contributions are of
minor importance, the dominating contributions come from the
two-particle twist-2 and twist-3 terms involving the $\phi_K(u)$,
$\phi_p(u)$ and $\phi_\sigma(u)$. We perform the same trick as
Refs.\cite{Belyaev94,Kim} and expand the amplitudes $\phi_K(u)$,
$\phi_p(u)$ and $\phi_\sigma(u)$ in terms of polynomials of $1-u$,
\begin{eqnarray}
\phi_K(u), \phi_p(u), \phi_\sigma(u),
\frac{d}{du}\phi_\sigma(u)&\Longrightarrow&\sum_{k=0}^N b_k(1-u)^k
\nonumber\\
 &=&\sum_{k=0}^N b_k \left(\frac{s-m_c^2}{s-q^2}\right)^k,
\end{eqnarray}
then introduce the variable $s'$ and the spectral densities  are
obtained. After straightforward but cumbersome calculations, we can
obtain the final expressions for the double Borel transformed
correlation functions $\Pi_\mu^{1(2)}$ at the level of quark-gluon
degrees of freedom below the thresholds. The masses of  the charmed
mesons are $M_{D^*}=2.012GeV$, $M_{D^*_s}=2.112GeV$,
$M_{D}=1.865GeV$, $M_{D_s}=1.97GeV$, $M_{D_0}=2.40GeV$,
 and $M_{D_{s0}}=2.317GeV$,
 the ratios  are
$\frac{M_D}{M_D+M_{D_{s0}}}\approx0.45$,
$\frac{M_D}{M_D+M_{D^*_s}}\approx0.47$,
$\frac{M_{D_s}}{M_{D_s}+M_{D^*}}\approx0.49$ and
$\frac{M_{D_s}}{M_{D_s}+M_{D_0}}\approx0.45$ \cite{PDG2004}.
 There exist overlapping working windows  for the two Borel
parameters $M_1^2$ and $M_2^2$. It's convenient to take the value
$M_1^2=M_2^2$,  $u_0=\frac{M_1^2}{M_1^2+M_2^2}=\frac{1}{2}$,
$M^2=\frac{M_1^2M_2^2}{M_1^2+M_2^2}=\frac{1}{2}M_1^2$,  furthermore,
the $K$ meson light-cone distribution amplitudes are known quite
well at the value $u_0=\frac {1}{2}$ comparing with the values at
the end-points.  We can introduce the threshold parameter $s_0$ and
make the simple replacement,
\begin{eqnarray}
\exp\left\{-\frac{m_c^2+u_0(1-u_0)m_K^2}{M^2}\right\} \rightarrow
\exp\left\{-\frac{m_c^2+u_0(1-u_0)m_K^2}{M^2}
\right\}-\exp\left\{-\frac{s_0}{M^2}\right\} \nonumber
\end{eqnarray}
 to subtract the contributions from the higher  resonances  and
  continuum states \cite{Belyaev94}, finally we obtain the following
  sum rules,

\begin{eqnarray}
 -\frac{ G_{D^* D_sK} m_{D^*}f_{D^*}f_{D_s}m_{D_s}^2}
  {m_c+m_s }\frac{m_{D^*}^2+m_{D_s}^2-m_K^2}{m_{D^*}^2}\exp\left\{-\frac{m_{D^*}^2}{M_1^2}-\frac{m_{D_s}^2}{M_2^2}\right\}
    =AA \, ;
\end{eqnarray}

\begin{eqnarray}
 -\frac{ G_{D^*_s DK} m_{D^*_s}f_{D^*_s}f_Dm_D^2}
  {m_c+m_u
  }\frac{m_{D^*_s}^2+m_D^2-m_K^2}{m_{D^*_s}^2}\exp\left\{-\frac{m_{D^*_s}^2}{M_1^2}-\frac{m_D^2}{M_2^2}\right\}
     =BB\, ;
\end{eqnarray}

\begin{eqnarray}
\frac{ G_{D^* D_sK} m_{D^*}f_{D^*}f_{D_s}m_{D_s}^2}
  {m_c+m_s } \frac{m_{D^*}^2-m_{D_s}^2+m_K^2}{m_{D^*}^2}\exp\left\{-\frac{m_{D^*}^2}{M_1^2}-\frac{m_{D_s}^2}{M_2^2}\right\}
  \nonumber\\+  \frac{ G_{D_0D_s K} f_{D_0}f_{D_s}m_{D_s}^2  }
  {m_c+m_s}\exp\left\{-\frac{m_{D_0}^2}{M_1^2}-\frac{m_{D_s}^2}{M_2^2}\right\}
  =CC\, ;
\end{eqnarray}

\begin{eqnarray}
\frac{ G_{D^*_s DK} m_{D^*_s}f_{D^*_s}f_D m_D^2}
  {m_c+m_u } \frac{m_{D^*_s}^2-m_D^2+m_K^2}{m_{D^*_s}^2}\exp\left\{-\frac{m_{D^*_s}^2}{M_1^2}-\frac{m_D^2}{M_2^2}\right\}
  \nonumber\\
   +\frac{ G_{D_{s0}D K} f_{D_{s0}}f_Dm_D^2  }
  {m_c+m_u}\exp\left\{-\frac{m_{D_{s0}}^2}{M_1^2}-\frac{m_D^2}{M_2^2}\right\}
  =DD\, ;
\end{eqnarray}

\begin{eqnarray}
 -\frac{ G_{D^* D_sK} m_{D^*}f_{D^*}f_{D_s}m_{D_s}^2}
  {m_c+m_s }\frac{m_{D^*}^2+m_{D_s}^2-m_K^2}{m_{D^*}^2}\exp\left\{-\frac{m_{D^*}^2}{M_1^2}-\frac{m_{D_s}^2}{M_2^2}\right\}
  \nonumber\\+  \frac{ G_{D_0D_s K} f_{D_0}f_{D_s}m_{D_s}^2  }
  { m_c+m_s }\exp\left\{-\frac{m_{D_0}^2}{M_1^2}-\frac{m_{D_s}^2}{M_2^2}\right\}
  =EE\, ;
\end{eqnarray}

\begin{eqnarray}
 -\frac{ G_{D^*_s D K} m_{D^*_s}f_{D^*_s}f_D m_D^2}
  {m_c+m_u
  }\frac{m_{D^*_s}^2+m_D^2-m_K^2}{m_{D^*_s}^2}\exp\left\{-\frac{m_{D^*_s}^2}{M_1^2}-\frac{m_D^2}{M_2^2}\right\}
  \nonumber \\
  +  \frac{ G_{D_{s0}D K} f_{D_{s0}}f_D m_D^2  }
  { m_c+m_u }\exp\left\{-\frac{m_{D_{s0}}^2}{M_1^2}-\frac{m_D^2}{M_2^2}\right\}
  =FF\, .
\end{eqnarray}
The explicit  expressions of the notations  $AA$, $BB$, $CC$, $DD$,
$EE$ and $FF$ are lengthy and given explicitly  in the appendix. A
slight different manipulation (with the techniques taken in the
Ref.\cite{ColangeloWang, WangW06}) for the dominating contributions
come from the terms involving the two-particle twist-2 and twist-3
light-cone distribution amplitudes $\phi_K(u)$, $\phi_p(u)$ and
$\phi_\sigma(u)$ leads to the sum rules with the same type as in
Ref. \cite{WangW06}. However, those type sum rules are not stable
with respect to the variations of the Borel parameter $M^2$, here we
will not show the expressions  explicitly   for simplicity. It is
not surprise that the QCD sum rules as a QCD model  have both
advantages and shortcomings.

\section{Numerical results and discussions}
The input parameters are taken as $m_s=(140\pm 10 )MeV$,
$m_c=(1.25\pm 0.10)GeV$, $\lambda_3=1.6\pm0.4$,
$f_{3K}=(0.45\pm0.15)\times 10^{-2}GeV^2$, $\omega_3=-1.2\pm0.7$,
$\omega_4=0.2\pm0.1$, $a_2=0.25\pm 0.15$, $a_1=0.06\pm 0.03$
\cite{LCSR,LCSRreview,Belyaev94,Ball98,Ball06}, $f_K=0.160GeV$,
$m_K=498MeV$, $m_{D_{s0}} =2.317GeV$, $m_D=1.865GeV$,
$m_{D_s}=1.97GeV$. In this article, we take the values of the $a_4$
to be zero,  and explore the dependence  of the strong coupling
constants $G_{D^*D_sK}$, $G_{D^*_sDK}$, $G_{D_0D_sK}$ and
$G_{D_{s0}DK}$ on  this parameter.

For the threshold parameter $s^0_{D_0}$, we can use the experimental
data as a guide, $m_{D_0}=2.40GeV$, $\Gamma_{m_{D_0}}=283MeV$
\cite{PDG2004}, and choose the values $s^0_{D_0}=(6.8-7.2)GeV^2$ to
subtract the contributions from the high resonances and continuum
states. The mass and width of the $D_0$ from Belle and Focus are
$m_{D_0}=2308\pm17\pm 15\pm 28 MeV$,
$\Gamma_{D_0}=276\pm21\pm18\pm60 MeV$ \cite{Belle04},
$m_{D_0}=2407\pm21\pm 35 MeV$, $\Gamma_{D_0}=240\pm55\pm59MeV$
\cite{Focus04}. The predictions from the constituent quark models
are $m_{D_0}=2.4GeV$  \cite{QuarkLattice}. The values of the mass
from the two collaborations have the difference about $100MeV$,  in
this article, we take the value $m_{D_0}=2.4GeV$ as input parameter,
our final numerical results for the large strong coupling constant
$G_{D_0D_sK}$ support smaller values for the $D_0$ if the same
mechanism takes place for both the charmed scalar  mesons $D_0$ and
$D_{s0}$. Furthermore, the strong coupling constant $G_{D_0D_sK}$ is
not sensitive to the values of the $m_{D_0}$, taking the values
$m_{D_0}=2.4GeV$ or $m_{D_0}=2.3GeV$ can not change the conclusion
qualitatively or quantitatively.

For the threshold parameters $s^0_{D^*}$, $s^0_{D_s^*}$ and
$s^0_{D_{s0}}$, the experimental values of the masses are
$m_{D^*}=2.01GeV$, $m_{D_s^*}=2.112GeV$ and $m_{D_{s0}}=2.317GeV$,
the widths are very narrow \cite{PDG2004}. We can  choose the values
of the threshold parameters $s^0_{D^*}=(4.7-5.1)GeV^2$,
$s^0_{D_s^*}=(4.8-5.2)GeV^2$ and $s^0_{D_{s0}}=(7.0-7.4)GeV^2$ to
subtract the contributions from the high resonances and continuum
states.  From Figs.1-3, we can see that the numerical values of the
strong coupling constants $G_{D^*D_sK}$ and $G_{D_0D_sK}$ are not
sensitive to the threshold parameters $s^0$ in those regions, the
values we chosen here are reasonable.
\begin{figure}
 \centering
 \includegraphics[totalheight=7cm]{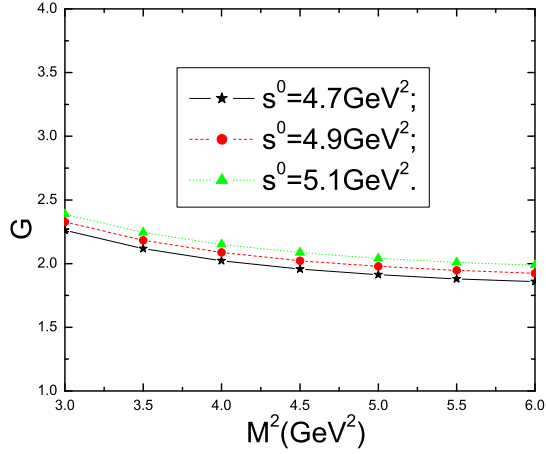}
 \caption{The   $G_{D^*D_{s}K}$ with the parameters $M^2$ and $s^0_{D^*}$ from Eq.(10). }
\end{figure}
\begin{figure}
 \centering
 \includegraphics[totalheight=7cm]{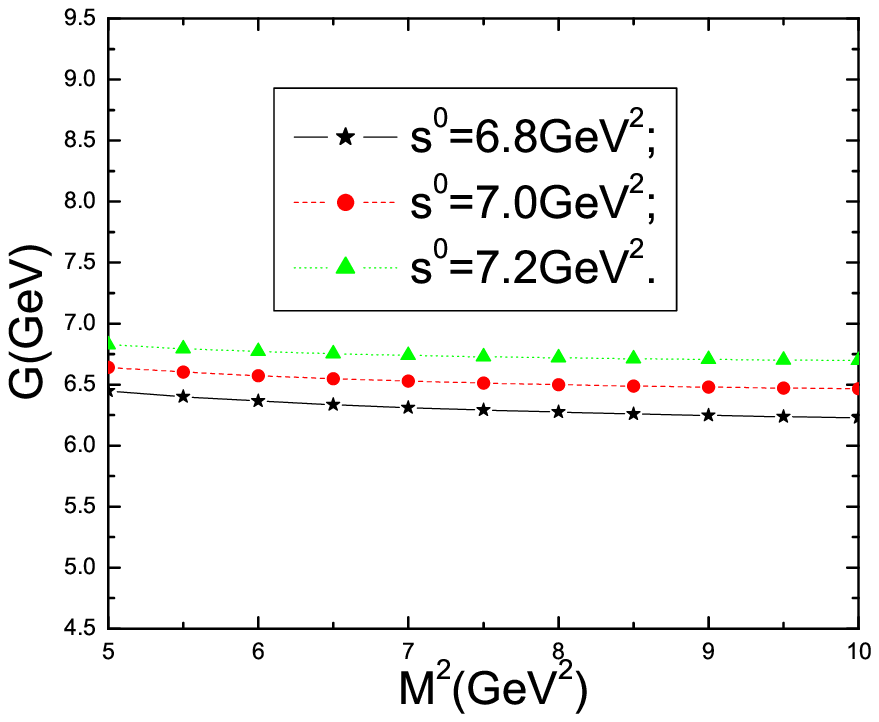}
 \caption{The   $G_{D_0D_{s}K}$ with the parameters $M^2$ and $s^0_{D_0}$ from Eq.(12). }
\end{figure}

\begin{figure}
 \centering
 \includegraphics[totalheight=7cm]{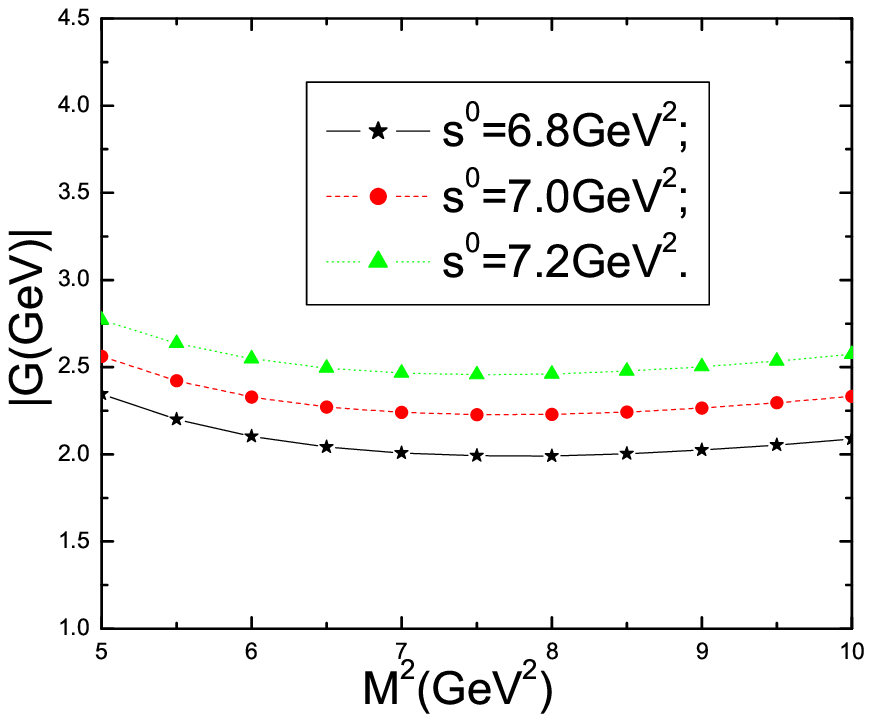}
 \caption{The   $G_{D_0D_{s}K}$ with the parameters $M^2$  and $s^0_{D_0}$ from Eq.(14). }
\end{figure}

The values of the decay constants $f_D$, $f_{D_s}$, $f_{D^*}$,
$f_{D^*_s}$, $f_{D_0}$ and $f_{D_{s0}}$  vary in a large range, for
example, $f_D=(0.17\pm 0.01)GeV$, $f_{D^*}=(0.24\pm 0.02)GeV$
\cite{Belyaev94}, $f_{D_0}=(0.217\pm 0.025) GeV$, $m_{D_0}=2.272GeV$
\cite{NarisonD03}, $f_{D_{s0}}=(0.225\pm0.025)GeV$
\cite{Colangelo2005}, $f_{D}=(0.177 \pm 0.021) GeV$, $ f_{D_{s}}=(0.
205 \pm 0.022) GeV$ \cite{BordesD05}, $f_{D_0}=(0.17\pm 0.02) GeV$
\cite{ColangeloD91} from the QCD sum rules; $f_{D_s}=0.268GeV$,
$f_{D^*_s}=0.315GeV$, $f_{D}=0.234GeV$, $f_{D^*}=0.310GeV$
\cite{EbertD06}, $f_{D^*_s}=0.375\pm 0.024 GeV$,  $f_{D^*}=0.340\pm
0.023 GeV$ \cite{GlwangD06}, $f_D=0.238GeV$, $f_{D_s}=0.241GeV$
\cite{WangD} from the potential models; $f_{D^*_s}=326^{+21}_{-17}
MeV$ , $f_{D^*}=223^{+23}_{-19} MeV$
 \cite{AlbertusD05} from the quark models, and $f_D=(222.6\pm16.7^{+2.8}_{-3.4})MeV$
 from the experimental data \cite{ExpD}. For a review of the  values of
 the decay constants for the mesons $D$ and $D_s$  from the  QCD sum rules and
 lattice QCD, one can consult the second article of the Ref.\cite{LCSRreview}.

 In this article, we take the following  constraints  for the decay constants,
\begin{eqnarray}
1.0<\frac{f_{D_{s0}}}{f_{D_0}}\approx
\frac{f_{D^*_{s}}}{f_{D^*}}\approx \frac{f_{D_{s}}}{f_D} < 1.1 \, ,
\end{eqnarray}
and choose the values,
\begin{eqnarray}
f_{D_s}=(0.25 \pm 0.02)GeV \, &,& \,f_{D}=(0.23 \pm 0.02)GeV   , \nonumber \\
f_{D_{s0}}=(0.225 \pm 0.025)GeV  \, &,& \, f_{D_0}=(0.217 \pm 0.020)GeV \, , \nonumber \\
f_{D^*_s}=(0.26 \pm 0.02)GeV \, &,& \, f_{D^*}=(0.24 \pm 0.02)GeV \,
\, .
\end{eqnarray}
In numerical calculation, we observe that the values of the strong
coupling constants $G_{D^*D_sK}$, $G_{D^*_sDK}$, $G_{D_0D_sK}$ and
$G_{D_{s0}DK}$ are sensitive to the six hadronic parameters, small
variations of those  parameters can lead to relatively large changes
for the numerical values, refining the six hadronic parameters is of
great importance.

The Borel parameters in Eqs.(10-11) are taken as $ M_1^2=M_2^2=
(6-12) GeV^2$ and $ M^2 =(3-6) GeV^2$, in those regions, the values
of the strong coupling constants $G_{D^*D_sK}$ and $G_{D^*_sDK}$ are
rather stable from the sum rules in Eqs.(10-11) with the simple
subtraction, which are shown, for example, in the Fig.1 and Figs.4-7
for the strong coupling constant $G_{D^*D_sK}$, similar figures can
be obtained if the values of the strong coupling constant
$G_{D^*_sDK}$ are plotted. In this article, we only show
   the numerical values from the sum rules in Eq.(10),
Eq.(12) and Eq.(14) explicitly for simplicity.

The Borel parameters in Eqs.(12-15) are chosen as $ M_1^2=M_2^2
=(10-20) GeV^2$ and $  M^2 =(5-10) GeV^2$, in those regions, the
values of  the strong coupling constants $G_{D_0D_sK}$ and
$G_{D_{s0}DK}$ are rather stable from the sum rules in Eqs.(12-13)
with the simple subtraction, which are shown  in the Fig2, Figs.4-6
and Fig.8 for an illustration. However, the   strong coupling
constants $G_{D_0D_sK}$ and $G_{D_{s0}DK}$ from the sum rules in
Eqs.(14-15) have a negative sign comparing with the corresponding
ones  from the sum rules in Eqs.(12-13), and much smaller  absolute
values. The fractions
\begin{eqnarray}
\frac{m_{D^*}^2+m_{D_s}^2-m_K^2}{m_{D^*}^2} \, , \,
\frac{m_{D^*_s}^2+m_D^2-m_K^2}{m_{D^*_s}^2} \nonumber
\end{eqnarray}
are about $2$. In the sum rules in Eqs.(10-11), the ground state
saturate condition can be safely satisfied below the threshold
$s^0_{D^*}$ ($s^0_{D^*_s}$). The vector interpolating current
$J_\mu^{D}(x)$ ($J_\mu^{D_s}(x)$)  has both non-vanishing couplings
 to the vector state $D^*$ ($D^*_s$) and to the scalar state $D_0$
($D_{s0}$), there are two hadronic states, the ground state $D^*$
($D^*_s$) and the first orbital excited state $D_0$ ($D_{s0}$) in
the channel $\bar{c}u$ ($\bar{c}s$) below the threshold $s^0_{D_0}$
($s^0_{D_{s0}}$), the ground states $D^*$ and $D^*_s$ are not
suppressed due to the factor $2$, the sum rules in Eqs.(14-15) are
not suitable for studying the strong coupling constants
$G_{D_0D_sK}$ and $G_{D_{s0}DK}$, our final numerical values support
this assumption. We show this fact in the Fig.3 for an illustration.

We determine the values of  the strong coupling constants
$G_{D^*D_sK}$ and $G_{D^*_sDK}$ from the Eq.(10) and Eq.(11)
respectively, then use those values as the input parameters, and
calculate the values of the strong coupling constants $G_{D_0D_sK}$
and $G_{D_{s0}DK}$ from the Eqs.(12-15) respectively.

The uncertainties of the five parameters  $\omega_4$, $\omega_3$,
$\lambda_3$, $m_c$ and $a_1$  can  not lead to large uncertainties
for the numerical values. The main uncertainties come from the ten
parameters $f_{3K}$, $m_s$,  $a_2$, $\eta_4$, $f_D$,  $f_{D^*}$,
$f_{D^*_s}$, $f_{D_0}$, $f_{D_s}$ and $f_{D_{s0}}$, small variations
of those parameters can lead to relatively large changes for the
numerical values, which are shown in the Figs.4-8 for an
illustration.
\begin{figure}
\centering
  \includegraphics[totalheight=7cm,width=7cm]{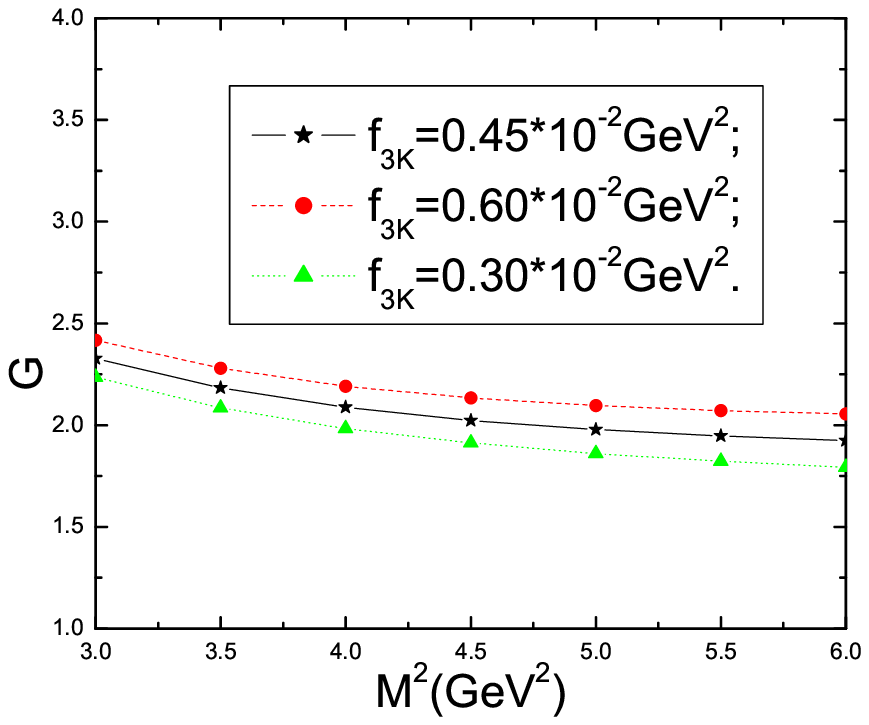}
  \includegraphics[totalheight=7cm,width=7cm]{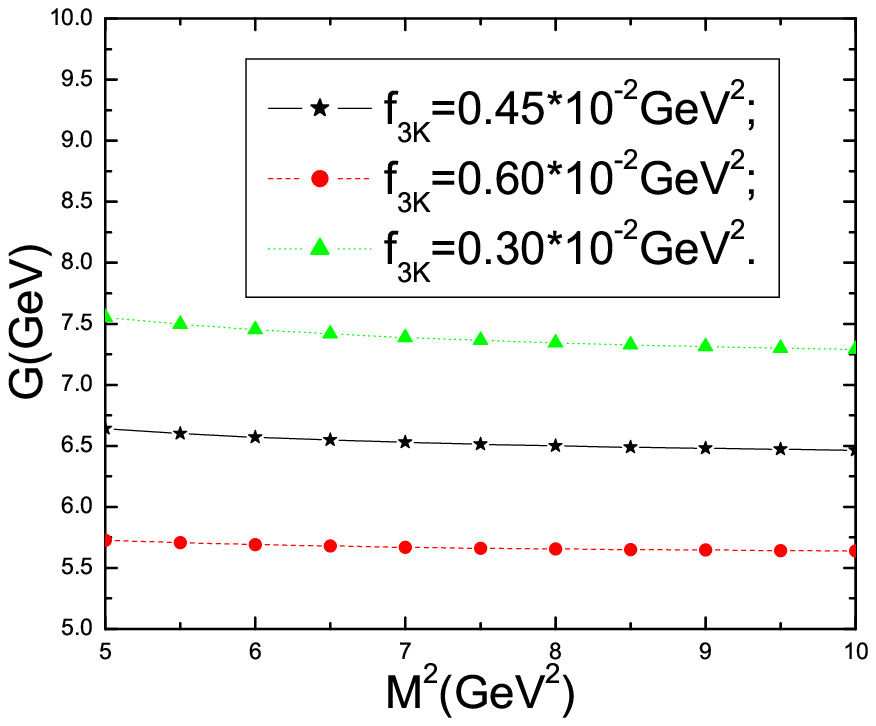}
 \caption{The   $G_{D^*D_{s}K}$ and $G_{D_0D_{s}K}$ with the parameters $M^2$ and $f_{3K}$ from Eq.(10) and
   Eq.(12) respectively. }
\end{figure}

\begin{figure}
 \centering
 \includegraphics[totalheight=7cm,width=7cm]{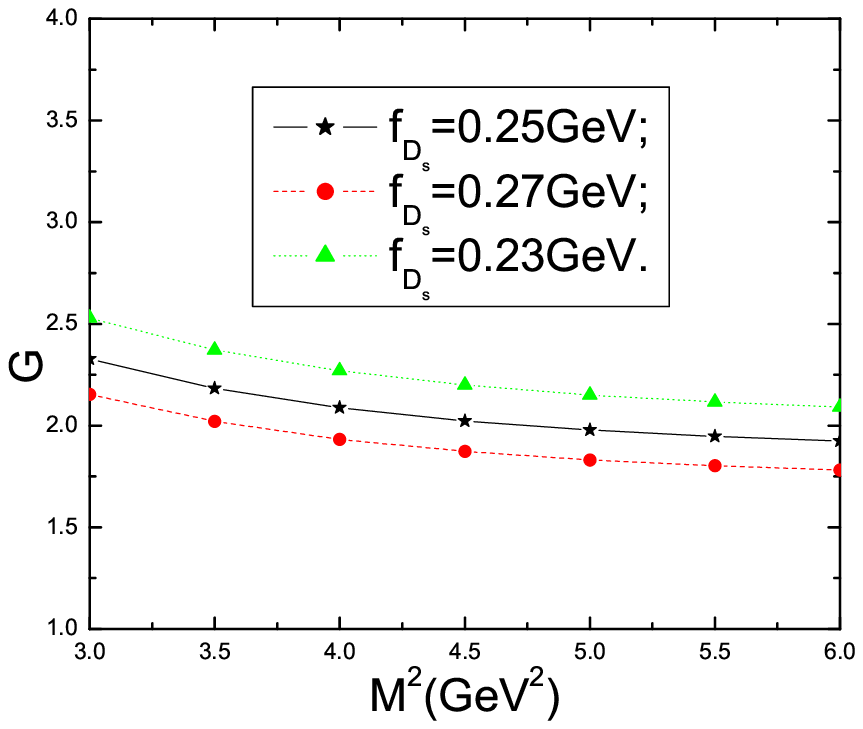}
 \includegraphics[totalheight=7cm,width=7cm]{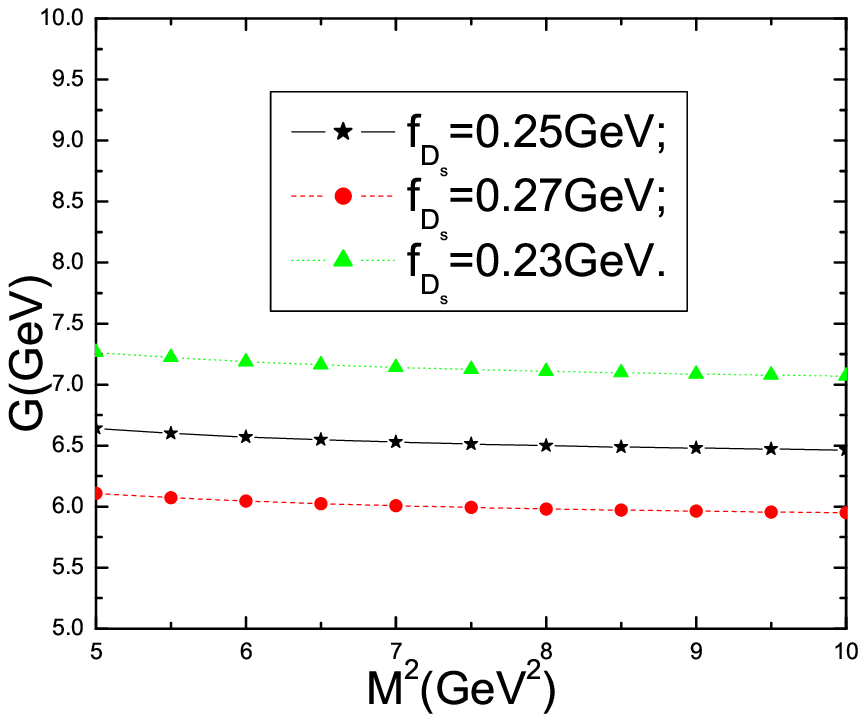}
 \caption{The   $G_{D^*D_{s}K}$ and $G_{D_0D_{s}K}$ with the parameters $M^2$ and $f_{D_s}$ from
 Eq.(10) and  Eq.(12) respectively.}
\end{figure}

\begin{figure}
 \centering
 \includegraphics[totalheight=7cm,width=7cm]{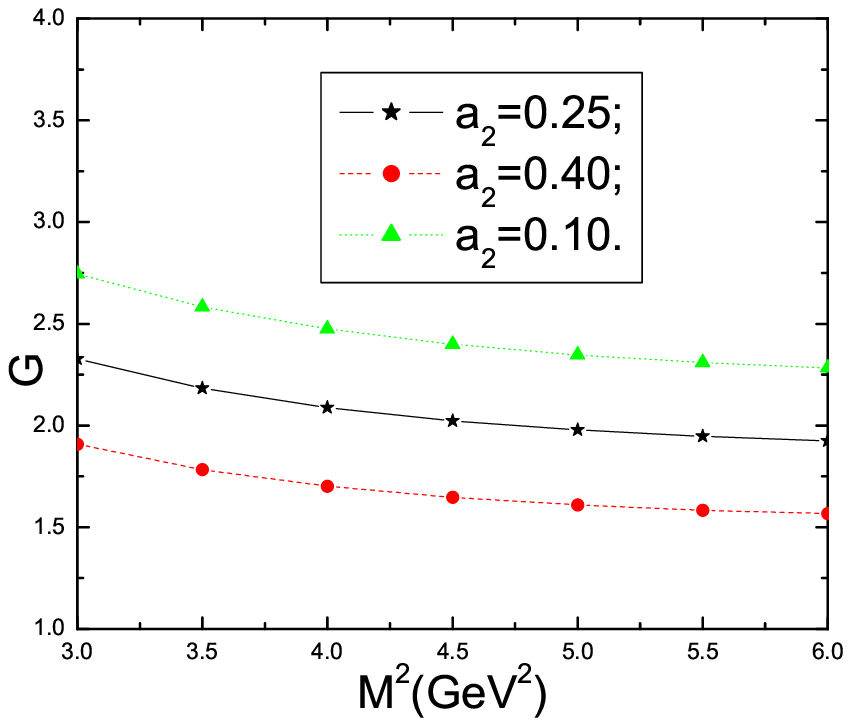}
 \includegraphics[totalheight=7cm,width=7cm]{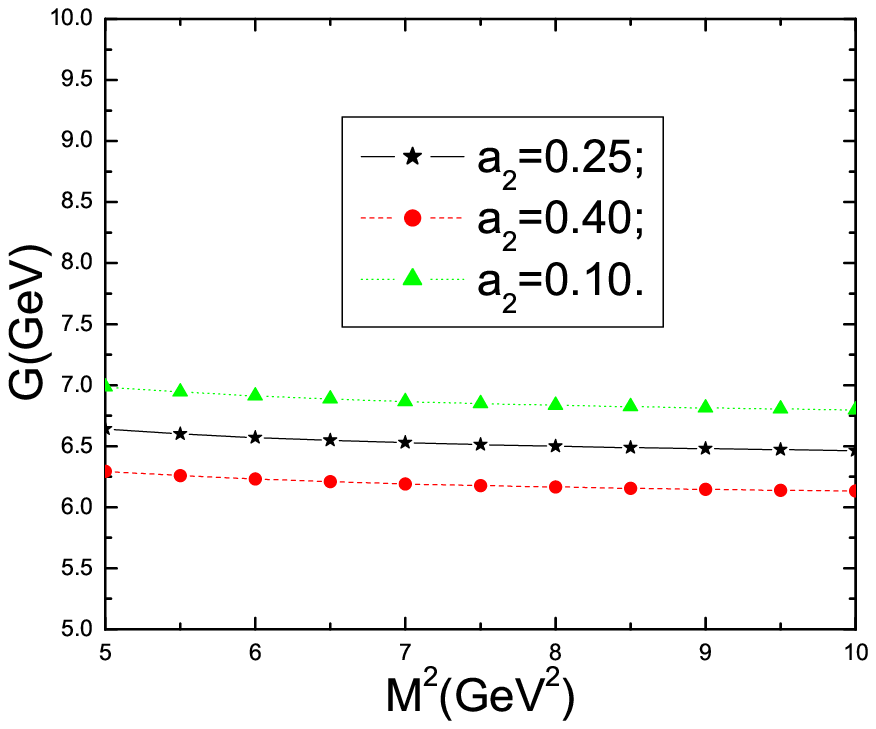}
 \caption{The   $G_{D^*D_{s}K}$ and $G_{D_0D_{s}K}$ with the parameters $M^2$ and $a_2$ from Eq.(10) and
  Eq.(12) respectively. }
\end{figure}

\begin{figure}
 \centering
 \includegraphics[totalheight=7cm,width=7cm]{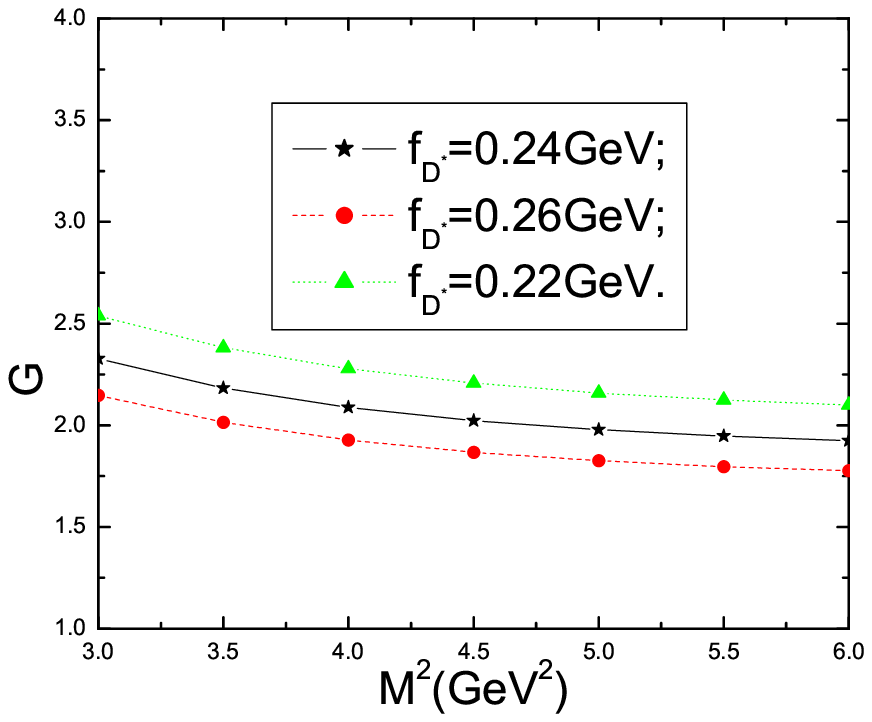}
  \caption{   The   $G_{D^*D_{s}K}$ with the parameters $M^2$ and $f_{D^*}$ from Eq.(10).
  }
\end{figure}

\begin{figure}
 \centering
  \includegraphics[totalheight=7cm,width=7cm]{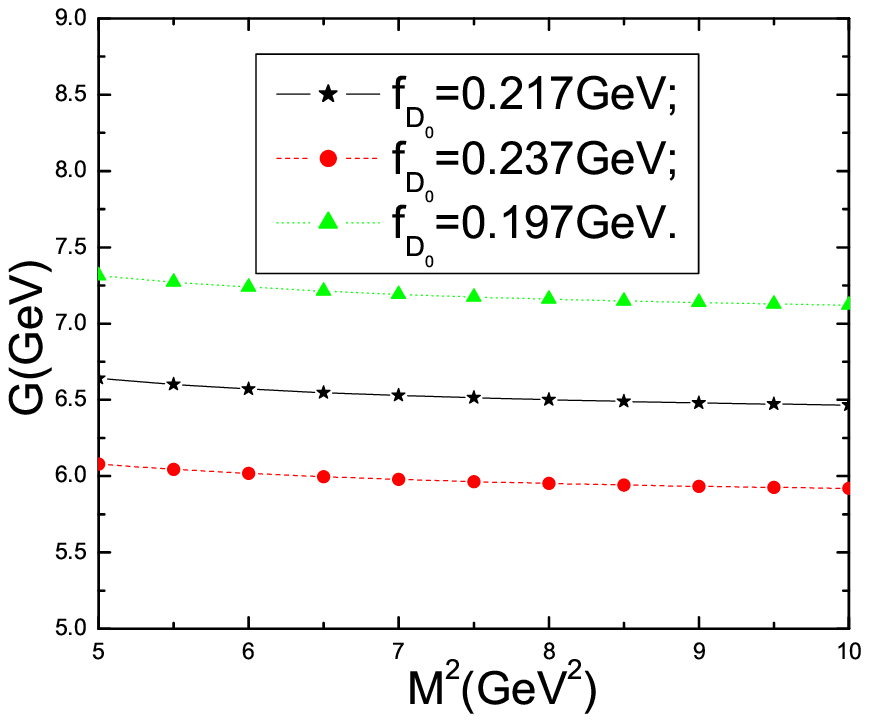}
 \includegraphics[totalheight=7cm,width=7cm]{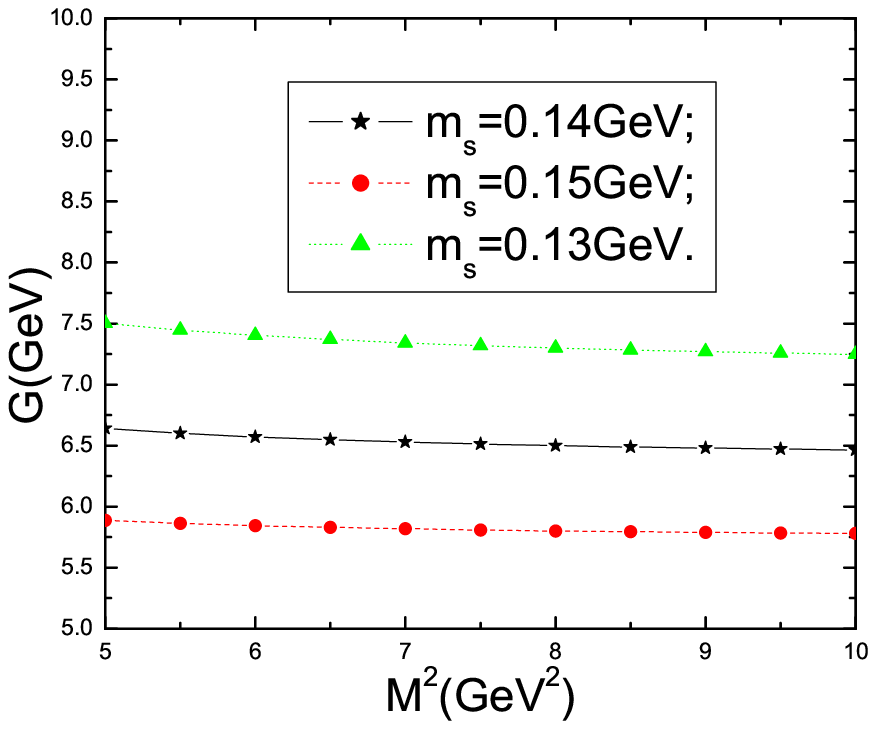}
  \includegraphics[totalheight=7cm,width=7cm]{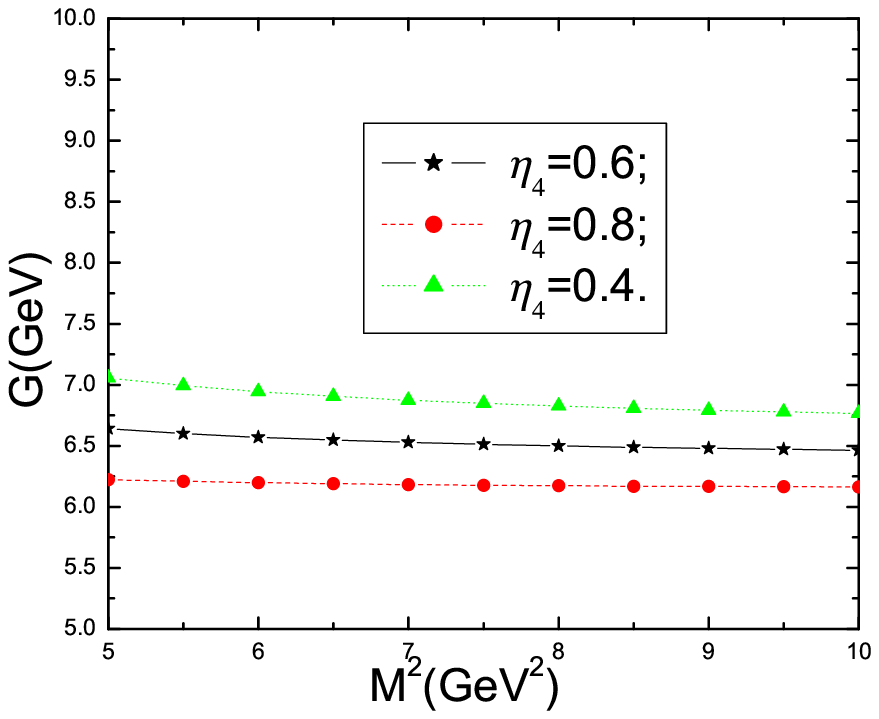}
 \caption{
 The   $G_{D_0D_{s}K}$ with the parameters $M^2$ and $f_{D_0}$, $m_s$, $\eta_4$ from Eq.(12). }
\end{figure}

Taking into account all the uncertainties, finally we obtain the
numerical results for the strong coupling constants,
\begin{eqnarray}
 G_{D^*D_sK} =2.02^{+0.84}_{-0.56}   \, , \, G_{D_0D_sK}
 =6.5^{+1.8}_{-1.5}
 GeV\, , \nonumber \\
 G_{D^*_sDK} =1.84^{+0.91}_{-0.63}   \, , \, G_{D_{s0}DK}
 =5.9^{+1.7}_{-1.6}
 GeV\, ,
\end{eqnarray}
which are shown in the Figs.9-10 respectively.
\begin{figure}
 \centering
 \includegraphics[totalheight=7cm,width=7cm]{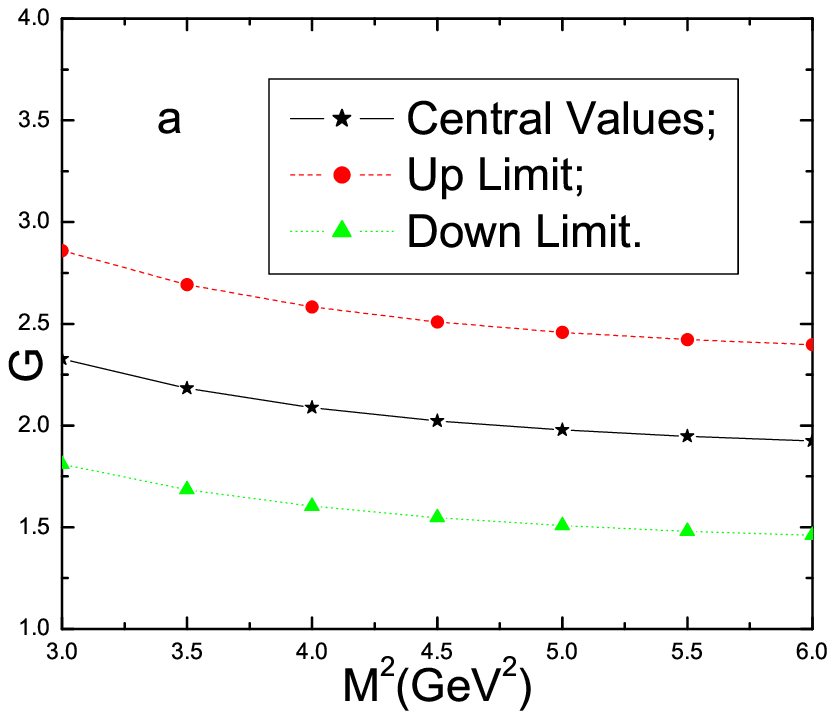}
 \includegraphics[totalheight=7cm,width=7cm]{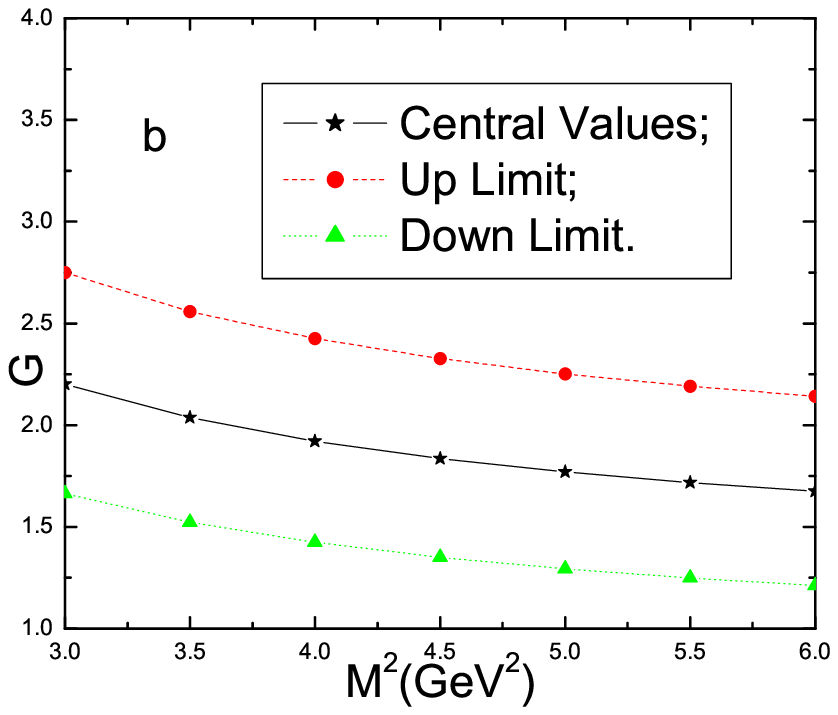}
 \caption{The   $G_{D^*D_{s}K}$(a) and $G_{D^*_sDK}$(b) with the parameter $M^2$ from Eq.(10) and Eq.(11) respectively.}
\end{figure}

\begin{figure}
 \centering
 \includegraphics[totalheight=7cm,width=7cm]{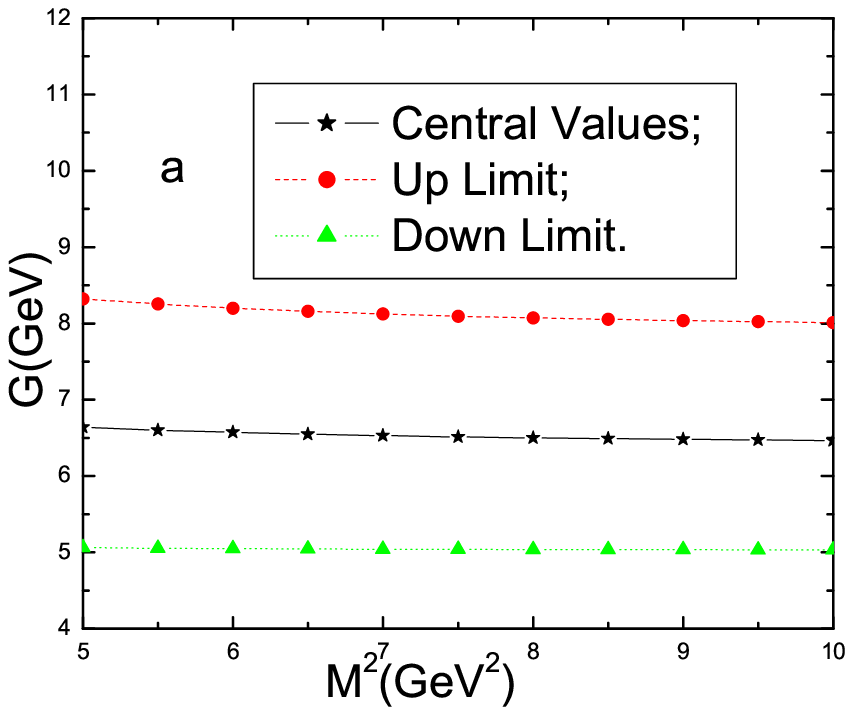}
 \includegraphics[totalheight=7cm,width=7cm]{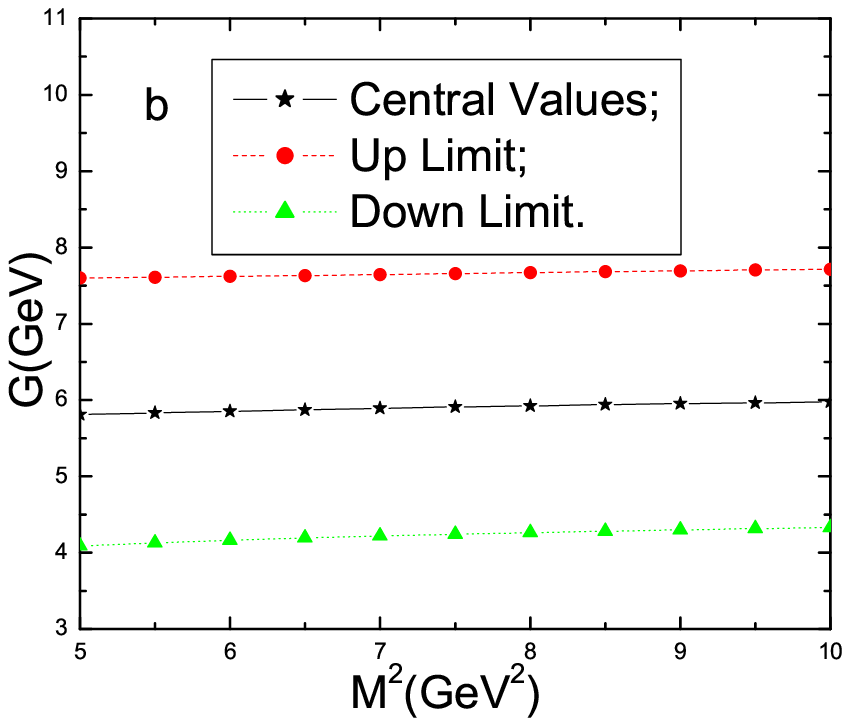}
 \caption{The   $G_{D_0D_{s}K}$(a) and $G_{D_{s0}DK}$(b) with the parameter $M^2$  from Eq.(12) and Eq.(13) respectively. }
\end{figure}

The strong coupling constants  $G_{D^*D_sK}$, $G_{D^*_sDK}$,
$G_{D_0D_sK}$ and $G_{D_{s0}DK}$ can be related to the parameters
$g$ and $h$ in the heavy-light Chiral perturbation theory
\cite{Casalbuoni97,Colangelo95},
\begin{eqnarray}
 G_{SP\pi} &=& \sqrt{m_{S} m_P} \frac{m_{S}^2-m_P^2}{m_{S}} \frac{|h|}{f_\pi} \,  , \nonumber \\
 G_{VP\pi} &=& \frac{2\sqrt{m_Pm_V}}{f_\pi}g \,  , \nonumber
\end{eqnarray}
here the $S$ are the  heavy  scalar  mesons with $0^+$, the $P$ are
the heavy pseudoscalar  mesons with $0^-$, the $V$ are the heavy
vector mesons with $1^-$, and  the $\pi$ stand for
 the light pseudoscalar mesons.

 The parameter $g$ has been calculated with the light-cone QCD sum
rules \cite{Colangelo97,Kim01,Khodjamirian99}, the quark models
\cite{Melikhov99,Becirevic99} and extracted from the experimental
data \cite{Colangelo02,Stewart98}. The values vary in a large range,
the corresponding values of  the strong coupling constants
$G_{D^*D_sK}$ and $G_{D^*_sDK}$ in the $SU(3)$ limit for the light
pseudoscalar mesons are listed in the Table.1. From the table, we
can see that our numerical results are compatible with the existing
estimations, although somewhat smaller.

The values of the strong coupling constants $G_{D^*D_sK}$ and
$G_{D^*_sDK}$ are sensitive to the non-perturbative parameter $a_4$,
if we take a larger value rather than zero,  larger values of the
$G_{D^*D_sK}$ and $G_{D^*_sDK}$ are obtained. The $G_{D^*D_{s}K}$
and $G_{D^*_sDK}$ are more sensitive to the $a_4$ comparing with the
$G_{D_0D_{s}K}$ and $G_{D_{s0}DK}$, which are shown in the Fig.11.
In fact, the largest uncertainties come from the uncertainties of
the $a_4$,   they are ideal channels to determine this parameter
directly from the experimental data. Once the experimental data for
the values of the strong coupling constants $G_{D^*D_{s}K}$ and
$G_{D^*_sDK}$ are available, powerful constraints can be put on the
range  of the parameter $a_4$. If we take the values from the QCD
sum rules as input parameters \cite{Bracco06},
$G_{D^*D_{s}K}=3.02\pm 0.14$ and $G_{D^*_sDK}=2.84\pm 0.31$, very
large values of  the $a_4$ are obtained.

 The parameter $h$ has been estimated
with the light-cone QCD sum rules \cite{Colangelo95}, the quark
models \cite{Becirevic99},  Adler-Weisberger type sum rules
\cite{Chow96}, and extracted from the experimental data
\cite{Mehen04}, the values are listed in the Table.2, from those
values we can estimate the values of the corresponding strong
coupling constants $G_{D_0D_sK}$ and $G_{D_{s0}DK}$ in the $SU(3)$
limit for the light pseudoscalar mesons. The value of the
dimensionless  effective coupling constant $\Gamma/k=0.46(9)$ from
Lattice QCD \cite{UKQCD04} is somewhat smaller than the values
extracted from the experimental data $\Gamma/k=0.73^{+28}_{-24}$,
here the $\Gamma$ is the decay width and the $k$ is the decay
momentum. Our numerical values $G_{D_0D_sK}=6.5^{+1.8}_{-1.5}GeV$
and $G_{D_{s0}DK}=5.9^{+1.7}_{-1.6}GeV$ are compatible with the
existing estimations in
Refs.\cite{Becirevic99,Colangelo95,Chow96,Mehen04}, although
somewhat smaller comparing with the values obtained in
Ref.\cite{WangW06} with the scalar interpolating current for  the
$D_{s0}$ meson, and about $2-3$ times as large as the energy scale
$M_{D_{s0}}=2.317GeV$, and favor the hadronic dressing mechanism.
For a short discussion about the hadronic dressing mechanism, one
can consult Ref.\cite{WangW06}, or one can consult the original
literatures for the details \cite{HDress,UQM}.

 The large values of  the strong coupling constants  $G_{D_0D_sK}$ and $G_{D_{s0}DK}$ obviously support
  the hadronic
dressing mechanism, the $D_0$ and $D_{s0}$ (just like the scalar
mesons $f_0(980)$ and $a_0(980)$, see Ref.\cite{ColangeloWang}) can
be taken as having small scalar $c\bar{u}$ and $c\bar{s}$ kernels of
typical meson size with large virtual S-wave $D_sK$ and $DK$ cloud
respectively. In Ref.\cite{Guo06}, the authors analyze the
unitarized two-meson scattering amplitudes from the heavy-light
Chiral Lagrangian,  and observe that the scalar meson $D_{s0}$
appears as the bound state pole with the strong coupling constant
$G_{D_{s0}DK}=10.203GeV$. Our numerical results $G_{D_{s0}DK} =
5.9^{+1.7}_{-1.6}  GeV$ are smaller, the values of our previously
work $G_{D_{s0}DK}=9.3^{+2.7}_{-2.1}GeV$ with the scalar
interpolating  current are more satisfactory \cite{WangW06}.

\begin{table}
\begin{center}
\begin{tabular}{c|c|c|c}
\hline\hline
      $|g|$ &$G_{D^*D_sK}$&$G_{D^*_sDK}$ &Reference  \\ \hline
      $0.38\pm0.08$& $9.5\pm2.0$&$9.4\pm2.0$ &\cite{Casalbuoni97}\\      \hline
     & $6.04\pm 0.28 $&$5.68\pm 0.62 $ &\cite{Bracco06}\\     \hline
    $0.34\pm0.10$& $8.5\pm2.5$ &$8.4\pm2.5$ &\cite{Colangelo97} \\ \hline
 0.28& 7.0& 6.9&\cite{Kim01} \\ \hline
          $0.35\pm0.10$& $8.7\pm2.5$ & $8.7\pm2.5$&\cite{Khodjamirian99} \\ \hline
  $0.50\pm0.02$& $12.4\pm0.5$ & $12.4\pm0.5$&\cite{Melikhov99} \\ \hline
  $0.61$&$15.2$ & $15.1$&\cite{Becirevic99} \\ \hline
         $0.59\pm0.07$& $14.7\pm1.7 $& $14.6\pm1.7$& \cite{Colangelo02}  \\ \hline
                                       $0.27^{+0.06}_{-0.03}$& $6.7^{+1.5}_{-0.7}$ &$6.7^{+1.5}_{-0.7}$ &\cite{Stewart98} \\ \hline
 $0.16^{+0.07}_{-0.05}$  &$4.04^{+1.68}_{-1.12}$&$3.68^{+1.82}_{-1.26}$&  This work
\\ \hline  \hline
\end{tabular}
\end{center}
\caption{ Numerical values of the  parameter $g$,  and the
corresponding values of the strong coupling constants $G_{D^*D_sK}$
and $G_{D^*_sDK}$ in the $SU(3)$ limit. Here we have double the
values of our numerical results and the ones from
Ref.\cite{Bracco06} due to the difference between the definitions
for the strong coupling constants. }
\end{table}

\begin{table}
\begin{center}
\begin{tabular}{c|c|c|c}
\hline\hline
      $|h|$& $G_{D_0D_sK}(GeV)$& $G_{D_{s0}DK}(GeV)$ &Reference  \\ \hline
$0.88^{+0.26}_{-0.20}$ &$9.4^{+2.8}_{-2.1}$ &$9.3^{+2.7}_{-2.1}$&
\cite{WangW06}\\ \hline
 &   & 10.203 &\cite{Guo06}\\ \hline
 0.536 &   $5.7$& $5.68$&\cite{Becirevic99} \\ \hline
       $0.52\pm0.17$&  $5.5\pm1.8$&  $5.5\pm1.8$& \cite{Colangelo95}  \\ \hline
                $ <0.93$&  $<9.9$& $<9.86$&\cite{Chow96} \\ \hline
        $0.57-0.74$&    6.1-7.9&6.0-7.8&  \cite{Mehen04}\\ \hline

                    $0.61^{+0.17}_{-0.14}$  (or $0.56^{+0.16}_{-0.15}$)     &    $6.5^{+1.8}_{-1.5}$&$5.9^{+1.7}_{-1.6}$& This work \\ \hline  \hline
\end{tabular}
\end{center}
\caption{ Numerical values of the  parameter $h$,  and the
corresponding values of the strong coupling constants $G_{D_0D_sK}$
and $G_{D_{s0}DK}$ in the $SU(3)$ limit. }
\end{table}

\begin{figure}
\centering
  \includegraphics[totalheight=7cm,width=7cm]{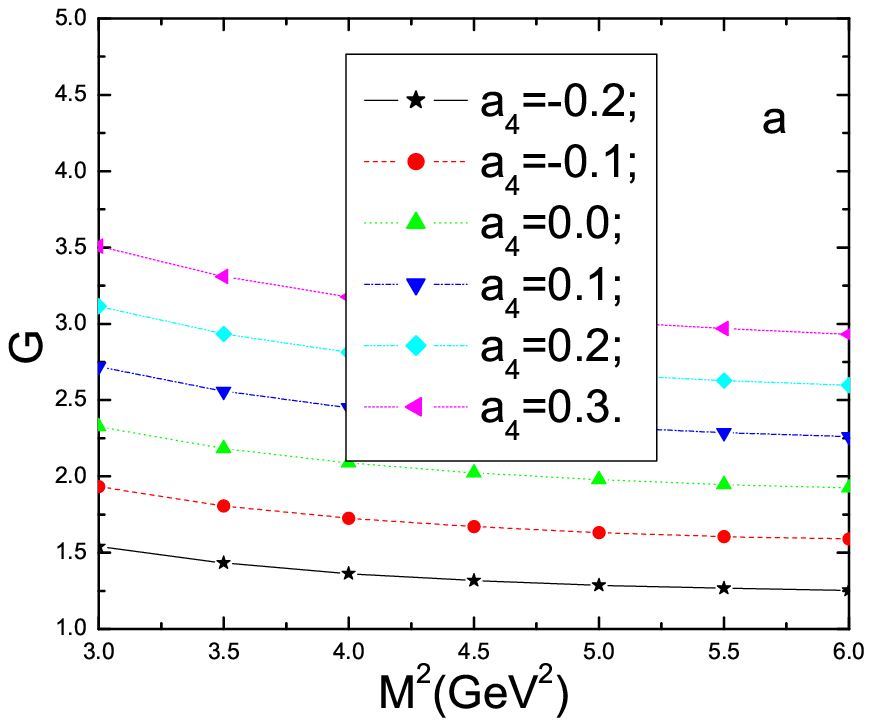}
  \includegraphics[totalheight=7cm,width=7cm]{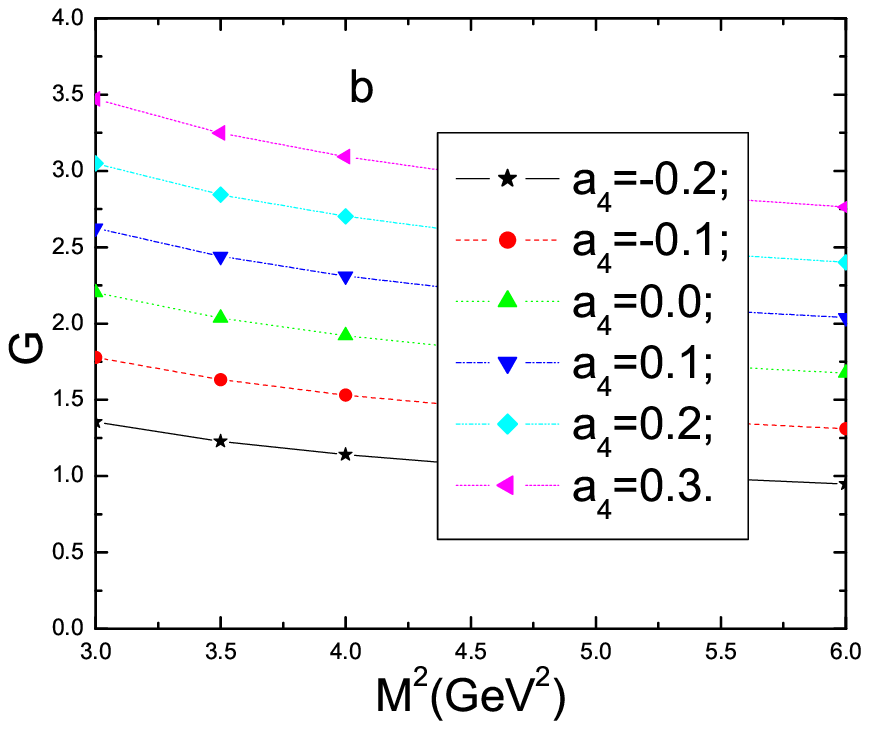}
  \includegraphics[totalheight=7cm,width=7cm]{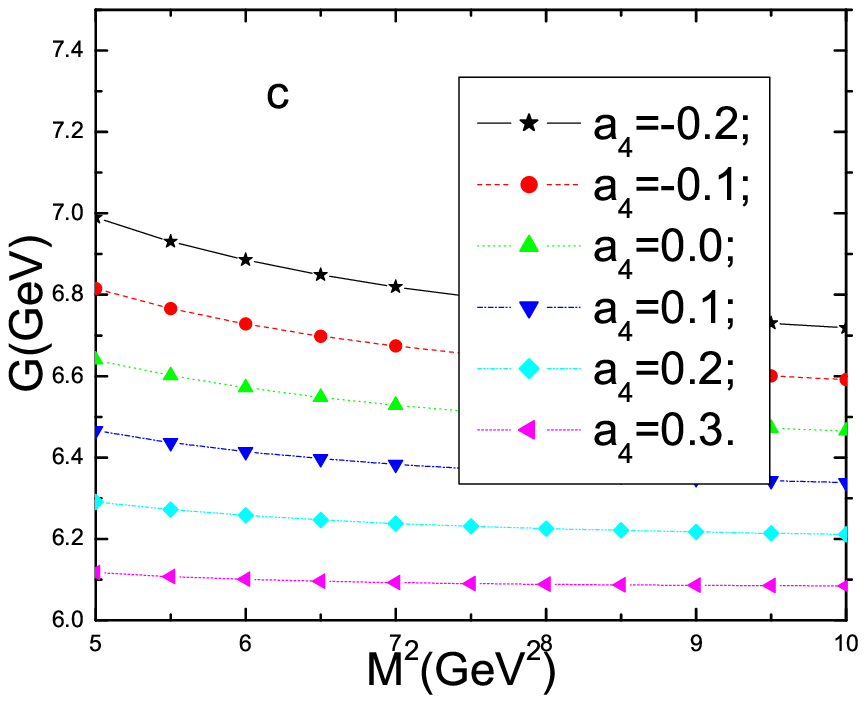}
\includegraphics[totalheight=7cm,width=7cm]{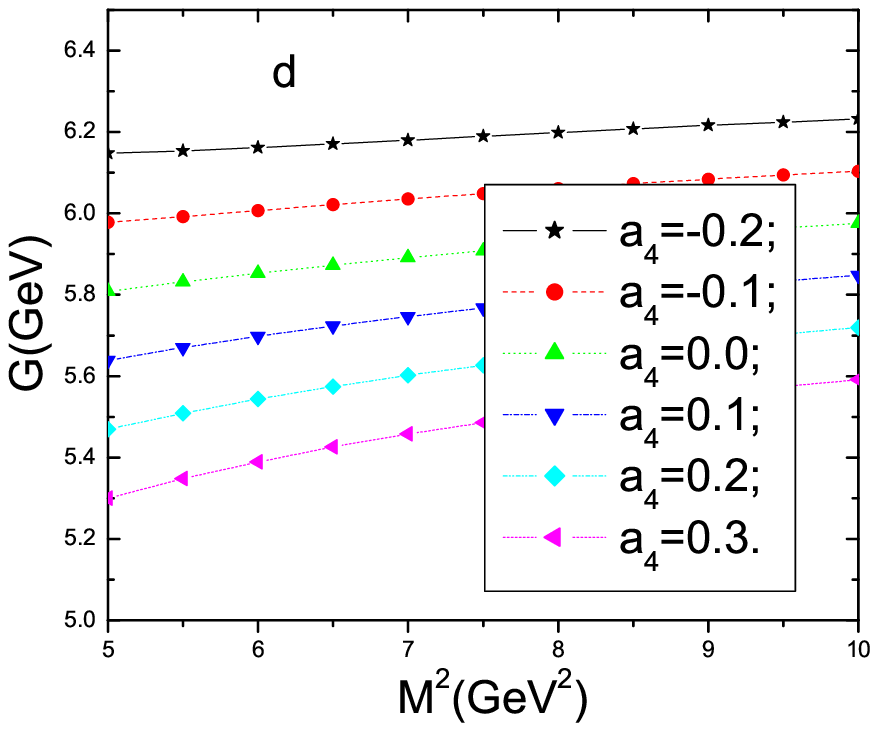}
 \caption{The   $G_{D^*D_{s}K}$(a), $G_{D^*_sDK}$(b), $G_{D_0D_{s}K}$(c), $G_{D_{s0}DK}$(d) with the parameters $M^2$  and $a_4$
 from Eq.(10),  Eq.(11),  Eq.(12),  Eq.(13) respectively . }
\end{figure}

\section{Conclusions}

In this article, we analyze the vertices $D^*D_sK$, $D^*_sDK$,
$D_0D_sK$ and $D_{s0}DK$ within the framework of the light-cone QCD
sum rules approach in an unified way. The strong coupling constants
$G_{D^*D_sK}$ and $G_{D^*_sDK}$ are important  parameters in
evaluating the charmonium absorption cross sections in searching for
the quark-gluon plasmas, our numerical values of the  $G_{D^*D_sK}$
and $G_{D^*_sDK}$ are compatible with the existing estimations
although somewhat smaller, the $SU(4)$ symmetry breaking effects are
very large, about $60\%$, the approximation of the $SU(4)$ symmetry
$G_{D^*D_sK}=G_{D^*_sDK}=5.0$ is not suitable \cite{SU4}. For the
scalar mesons $D_0$ and $D_{s0}$, we  take the point of view that
they are the conventional $c\bar{u}$ and $c\bar{s}$ meson
respectively, and calculate the strong coupling constants $G_{D_0
D_s K}$ and $G_{D_{s0} D K}$ within the framework of the light-cone
QCD sum rules approach. The numerical values of the scalar-$D_sK$
and -$DK$ coupling constants $G_{D_0 D_s K}$ and $G_{D_{s0} D K}$
are compatible with the existing estimations although somewhat
smaller, the large values support the hadronic dressing mechanism.
Just like the scalar mesons $f_0(980)$ and $a_0(980)$, the scalar
mesons $D_0$ and $D_{s0}$ may have small $c\bar{u}$ and  $c\bar{s}$
kernels  of typical $c\bar{u}$ and $c\bar{s}$ mesons  size
respectively. The strong coupling to virtual intermediate hadronic
states (or the virtual mesons loops) can result in  smaller mass
than the conventional scalar  mesons $c\bar{u}$ and $c\bar{s}$ in
the constituent quark models, enrich the pure states $c\bar{u}$ and
$c\bar{s}$   with other components. The $D_0 $  and  $D_{s0} $ may
spend part (or most part) of they lifetime as virtual $ D_s K $ and
$ D K $ states. Furthermore, we study the dependence   of the strong
coupling constants $G_{D^*D_sK}$ and $G_{D^*_sDK}$ on the
non-perturbative parameter $a_4$ of the twist-2 $K$ meson light-cone
distribution amplitude. The values of the strong coupling constants
$G_{D^*D_{s}K}$ and $G_{D^*_sDK}$ are more sensitive to the  $a_4$
comparing with the $G_{D_0D_{s}K}$ and $G_{D_{s0}DK}$. The largest
uncertainties come from the uncertainties of the $a_4$, they are the
ideal channels to determine the parameter directly from the
experimental data. Once the experimental data for the values of the
strong coupling constants $G_{D^*D_{s}K}$ and $G_{D^*_sDK}$  are
available, powerful constraints can be put on the range  of the
parameter $a_4$.

 \section*{Appendix}

The explicit expressions of the correlation functions $\Pi^1_\mu$
and $\Pi^2_\mu$ in the hadronic representation,
\begin{eqnarray}
\Pi^1_\mu&=&\frac{<0\mid J^D_{\mu}\mid D^*(q+P)><D^* D_s\mid K>
<D_{s}(q)|J_{D_{s}}\mid 0>}
  {\left\{m_{D^*}^2-(q+P)^2\right\}\left(m_{D_{s}}^2-q^2\right)} \nonumber \\
  &&+\frac{<0\mid J^D_{\mu}\mid D_0(q+P)><D_0 D_s\mid K>
<D_{s}(q)|J_{D_{s}}\mid 0>}
  {\left\{m_{D_0}^2-(q+P)^2\right\}\left(m_{D_{s}}^2-q^2\right)}+\cdots \nonumber \\
  &=&\frac{ G_{D^*D_s K} m_{D^*}f_{D^*}f_{D_{s}}m_{D_{s}}^2  (P-q)\cdot \epsilon \epsilon_\mu}
  {\left(m_c+m_s\right) \left\{m_{D^*}^2-(q+P)^2\right\}\left(m_{D_s}^2-q^2\right)} \nonumber\\
&&+\frac{ G_{D_0D_s K} f_{D_0}f_{D_s}m_{D_s}^2  (q+P)_\mu}
  {\left(m_c+m_s\right)\left\{m_{D_0}^2-(q+P)^2\right\}\left(m_{D_s}^2-q^2\right)} +
  \cdots\nonumber \\
  &=&\left\{\frac{ G_{D^* D_sK} m_{D^*}f_{D^*}f_{D_s}m_{D_s}^2}
  {\left(m_c+m_s\right) \left\{m_{D^*}^2-(q+P)^2\right\}\left(m_{D_s}^2-q^2\right)} \frac{m_{D^*}^2-m_{D_s}^2+m_K^2}{m_{D^*}^2}\right. \nonumber\\
  &&\left.+  \frac{ G_{D_0D_s K} f_{D_0}f_{D_s}m_{D_s}^2  }
  {\left(m_c+m_s\right)\left\{m_{D_{0}}^2-(q+P)^2\right\}\left(m_{D_s}^2-q^2\right)}\right\}q_\mu+ \nonumber \\
&&+ \left\{-\frac{ G_{D^* D_sK} m_{D^*}f_{D^*}f_{D_s}m_{D_s}^2}
  {\left(m_c+m_s\right) \left\{m_{D^*}^2-(q+P)^2\right\}\left(m_{D_s}^2-q^2\right)} \frac{m_{D^*}^2+m_{D_s}^2-m_K^2}{m_{D^*}^2}\right. \nonumber\\
  &&\left.+  \frac{ G_{D_0D_s K} f_{D_0}f_{D_s}m_{D_s}^2  }
  {\left(m_c+m_s\right)\left\{m_{D_{0}}^2-(q+P)^2\right\}\left(m_{D_s}^2-q^2\right)}\right\}P_\mu+ \cdots  ,
\end{eqnarray}

\begin{eqnarray}
\Pi^2_\mu&=&\frac{<0\mid J^{D_s}_{\mu}\mid D_s^*(q+P)><D_s^* D\mid
K> <D(q)|J_{D}\mid 0>}
  {\left\{m_{D_s^*}^2-(q+P)^2\right\}\left(m_{D}^2-q^2\right)} \nonumber \\
  &&+\frac{<0\mid J^{D_s}_{\mu}\mid D_{s0}(q+P)><D_{s0} D\mid K>
<D(q)|J_{D}\mid 0>}
  {\left\{m_{D_{s0}}^2-(q+P)^2\right\}\left(m_{D}^2-q^2\right)}+\cdots \nonumber \\
  &=&\frac{ G_{D_s^*DK} m_{D^*_s}f_{D^*_s}f_{D}m_D^2  (P-q)\cdot \epsilon \epsilon_\mu}
  {\left(m_c+m_u\right) \left\{m_{D_s^*}^2-(q+P)^2\right\} \left(m_D^2-q^2\right)} \nonumber\\
&&+\frac{ G_{D_{s0}DK} f_{D_{s0}}f_D m_D^2  (q+P)_\mu}
  {\left(m_c+m_u\right) \left\{m_{D_{s0}}^2-(q+P)^2\right\}  \left(m_D^2-q^2\right)} +
  \cdots\nonumber \\
    &=&\left\{\frac{ G_{D^*_{s}DK} m_{D^*_s}f_{D^*_s}f_D m_D^2}
  {\left(m_c+m_u\right)\left\{m_{D^*_s}^2-(q+P)^2\right\} \left(m_D^2-q^2\right)}\frac{m_{D^*_s}^2-m_D^2+m_K^2}{m_{D^*_s}^2}\right. \nonumber\\
  &&+\left.  \frac{ G_{D_{s0}DK} f_{D_{s0}}f_{D}m_D^2  }
  {\left(m_c+m_u\right)\left\{m_{D_{s0}}^2-(q+P)^2\right\}\left(m_D^2-q^2\right)} \right\}q_\mu +\nonumber\\
&&\left\{-\frac{ G_{D^*_{s}DK} m_{D^*_s}f_{D^*_s}f_D m_D^2}
  {\left(m_c+m_u\right)\left\{m_{D^*_s}^2-(q+P)^2\right\} \left(m_D^2-q^2\right)}\frac{m_{D^*_s}^2+m_D^2-m_K^2}{m_{D^*_s}^2}\right. \nonumber\\
  &&+\left. \frac{ G_{D_{s0}DK} f_{D_{s0}}f_{D}m_D^2  }
  {\left(m_c+m_u\right)\left\{m_{D_{s0}}^2-(q+P)^2\right\}\left(m_D^2-q^2\right)} \right\}P_\mu + \cdots
  .
\end{eqnarray}

The explicit expressions of the correlation functions $\Pi^1_\mu$
and $\Pi^2_\mu$ at the level of quark-gluon degrees of freedom,
\begin{eqnarray}
\Pi_\mu^1&=&q_\mu\left\{ -\frac{f_Km_K^2}{m_s}\int_0^1du
\frac{1}{(q+uP)^2-m_c^2}
\left[\phi_p(u)-\frac{d}{6du}\phi_\sigma(u)\right]\right. \nonumber\\
&&+m_cf_Km_K^2\int_0^1du\int_0^udt\frac{B(t)}{\left\{(q+uP)^2-m_c^2\right\}^2}
\nonumber\\
&&+f_{3K}m_K^2\int_0^1dv \int_0^1d\alpha_g
\int_0^{1-\alpha_g}d\alpha_s
\frac{(2v-3)T(\alpha_u,\alpha_g,\alpha_s)}{\left\{\left[q+((1-v)\alpha_g+\alpha_s)P\right]^2-m_c^2\right\}^2}
\nonumber\\
&&-4m_cf_Km_K^4\int_0^1dv v \int_0^1 d\alpha_g\int_0^{\alpha_g}
d\beta\int_0^{1-\beta}d\alpha
\frac{\Phi(1-\alpha-\beta,\beta,\alpha)}{\left\{\left[q+(1-v\alpha_g)P\right]^2-m_c^2\right\}^3}
\nonumber \\
&&\left.+4m_cf_Km_K^4\int_0^1 dv\int_0^1
d\alpha_g\int_0^{1-\alpha_g} d\alpha_s
 \int_0^{\alpha_s}d\alpha
\frac{\Phi(1-\alpha-\alpha_g,\alpha_g,\alpha)}{\left\{\left[q+((1-v)\alpha_g+\alpha_s)P\right]^2-m_c^2\right\}^3}\right\}
\nonumber
\end{eqnarray}
\begin{eqnarray}
&+&P_\mu\left\{-\frac{f_Km_K^2}{m_s}\int_0^1du
\frac{u\phi_p(u)}{(q+uP)^2-m_c^2} +m_cf_Km_K^2\int_0^1du\int_0^udt\frac{uB(t)}{\left\{(q+uP)^2-m_c^2\right\}^2}\right. \nonumber \\
&&+\frac{f_Km_K^2}{6m_s}\int_0^1\phi_\sigma(u)\left\{
\left[1-u\frac{d}{du}\right]\frac{1}{(q+uP)^2-m_c^2}-\frac{2m_c^2}{\left[(q+uP)^2-m_c^2\right]^2}
\right\} \nonumber\\
&&+m_cf_K\int_0^1du \left\{
\frac{\phi_K}{(q+uP)^2-m_c^2}-\frac{m_K^2m_c^2}{2}\frac{A(u)}{\left[(q+uP)^2-m_c^2\right]^3}
\right\} \nonumber\\
&&+f_{3K}m_K^2 \int_0^1dv\int_0^1d\alpha_g
\int_0^{1-\alpha_g}d\alpha_s \frac{
\left[(1-v)\alpha_g+\alpha_s\right]\left(2v
-3\right)T(\alpha_u,\alpha_g,\alpha_s)}
{\left\{\left[q+((1-v)\alpha_g+\alpha_s)P\right]^2-m_c^2\right\}^2}\nonumber\\
&&-2f_{3K} \int_0^1dvv \int_0^1d\alpha_g
\int_0^{1-\alpha_g}d\alpha_s
T(\alpha_u,\alpha_g,\alpha_s)\frac{d}{du}\frac{1}{
(q+uP)^2-m_c^2}\mid_{u=(1-v)\alpha_g+\alpha_s}\nonumber
\\
 &&-4m_cf_Km_K^4\int_0^1dvv \int_0^1 d\alpha_g\int_0^{\alpha_g}
d\beta\int_0^{1-\beta}d\alpha
\frac{(1-v\alpha_g)\Phi(1-\alpha-\beta,\beta,\alpha)}{\left\{\left[q+(1-v\alpha_g)P\right]^2-m_c^2\right\}^3}
\nonumber \\
&&+4m_cf_Km_K^4\int_0^1 dv\int_0^1 d\alpha_g\int_0^{1-\alpha_g}
d\alpha_s
 \int_0^{\alpha_s}d\alpha \frac{((1-v)\alpha_g+\alpha_s)\Phi(1-\alpha-\alpha_g,\alpha_g,\alpha)}
{\left\{\left[q+((1-v)\alpha_g+\alpha_s)P\right]^2-m_c^2\right\}^3}
\nonumber \\
&&\left. -m_cf_Km_K^2\int_0^1 dv\int_0^1
d\alpha_g\int_0^{1-\alpha_g} d\alpha_s
\frac{\Psi(\alpha_u,\alpha_g,\alpha_s)}
{\left\{\left[q+((1-v)\alpha_g+\alpha_s)P\right]^2-m_c^2\right\}^2}\right\}
\, ,
\end{eqnarray}
\begin{eqnarray}
\Pi^2_\mu=\Pi^1_\mu\left( u\longleftrightarrow 1-u; \alpha_u
\longleftrightarrow \alpha_s \right) \, .
\end{eqnarray}

The light-cone distribution amplitudes of the $K$ meson,
\begin{eqnarray}
<0| {\bar u} (0) \gamma_\mu \gamma_5 s(x) |K(P)>& =& i f_K P_\mu
\int_0^1 du  e^{-i u P\cdot x}
\left\{\varphi_K(u)+\frac{m_K^2x^2}{16}
A(u)\right\}\nonumber\\
&&+f_K m_K^2\frac{ix_\mu}{2P\cdot x}
\int_0^1 du  e^{-i u P \cdot x} B(u) \, , \nonumber\\
<0| {\bar u} (0) i \gamma_5 s(x) |K(P)> &=& \frac{f_K M_K^2}{ m_s}
\int_0^1 du  e^{-i u P \cdot x} \varphi_p(u)  \, ,  \nonumber\\
<0| {\bar u} (0) \sigma_{\mu \nu} \gamma_5 s(x) |K(P)> &=&i(P_\mu
x_\nu-P_\nu x_\mu)  \frac{f_K M_K^2}{6 m_s} \int_0^1 du
e^{-i u P \cdot x} \varphi_\sigma(u) \, ,  \nonumber\\
<0| {\bar u} (0) \sigma_{\alpha \beta} \gamma_5 g_s G_{\mu \nu}(v
x)s(x) |K(P)>&=& f_{3 K}\left\{(P_\mu P_\alpha g^\bot_{\nu
\beta}-P_\nu P_\alpha g^\bot_{\mu \beta}) -(P_\mu P_\beta g^\bot_{\nu \alpha}\right.\nonumber\\
&&\left.-P_\nu P_\beta g^\bot_{\mu \alpha})\right\} \int {\cal
D}\alpha_i \varphi_{3 K} (\alpha_i)
e^{-iP \cdot x(\alpha_s+v \alpha_g)} \, ,\nonumber\\
<0| {\bar u} (0) \gamma_{\mu} \gamma_5 g_s G_{\alpha \beta}(vx)s(x)
|K(P)>&=&  P_\mu  \frac{P_\alpha x_\beta-P_\beta x_\alpha}{P
\cdot x}f_Km_K^2\nonumber\\
&&\int{\cal D}\alpha_i A_{\parallel}(\alpha_i) e^{-iP\cdot
x(\alpha_s +v \alpha_g)}\nonumber \\
&&+ f_Km_K^2 (P_\beta g_{\alpha\mu}-P_\alpha
g_{\beta\mu})\nonumber\\
&&\int{\cal D}\alpha_i A_{\perp}(\alpha_i)
e^{-iP\cdot x(\alpha_s +v \alpha_g)} \, ,  \nonumber\\
<0| {\bar u} (0) \gamma_{\mu}  g_s \tilde G_{\alpha \beta}(vx)s(x)
|K(P)>&=& P_\mu  \frac{P_\alpha x_\beta-P_\beta x_\alpha}{P \cdot
x}f_Km_K^2\nonumber\\
&&\int{\cal D}\alpha_i V_{\parallel}(\alpha_i) e^{-iP\cdot
x(\alpha_s +v \alpha_g)}\nonumber \\
&&+ f_Km_K^2 (P_\beta g_{\alpha\mu}-P_\alpha
g_{\beta\mu})\nonumber\\
&&\int{\cal D}\alpha_i V_{\perp}(\alpha_i) e^{-iP\cdot x(\alpha_s +v
\alpha_g)} \, ,
\end{eqnarray}
here the operator $\tilde G_{\alpha \beta}$  is the dual of the
$G_{\alpha \beta}$, $\tilde G_{\alpha \beta}= {1\over 2}
\epsilon_{\alpha \beta  \mu\nu} G^{\mu\nu} $, ${\cal{D}}\alpha_i$ is
defined as ${\cal{D}} \alpha_i =d \alpha_1 d \alpha_2 d \alpha_3
\delta(1-\alpha_1 -\alpha_2 -\alpha_3)$,
$\Phi(\alpha_1,\alpha_2,\alpha_3)=A_\perp+V_\perp+A_\parallel+V_\parallel$
and
$\Psi(\alpha_1,\alpha_2,\alpha_3)=A_\parallel+V_\parallel-2A_\perp-2V_\perp$.
The  light-cone distribution amplitudes are parameterized as
\begin{eqnarray}
\phi_K(u,\mu)&=&6u(1-u)
\left\{1+a_1C^{\frac{3}{2}}_1(2u-1)+a_2C^{\frac{3}{2}}_2(2u-1)
+a_4C^{\frac{3}{2}}_4(2u-1)\right\}\, , \nonumber\\
\varphi_p(u,\mu)&=&1+\left\{30\eta_3-\frac{5}{2}\rho^2\right\}C_2^{\frac{1}{2}}(2u-1)\nonumber \\
&&+\left\{-3\eta_3\omega_3-\frac{27}{20}\rho^2-\frac{81}{10}\rho^2 a_2\right\}C_4^{\frac{1}{2}}(2u-1)\, ,  \nonumber \\
\varphi_\sigma(u,\mu)&=&6u(1-u)\left\{1
+\left[5\eta_3-\frac{1}{2}\eta_3\omega_3-\frac{7}{20}\rho^2-\frac{3}{5}\rho^2 a_2\right]C_2^{\frac{3}{2}}(2u-1)\right\}\, , \nonumber \\
T(\alpha_i,\mu) &=& 360 \alpha_u \alpha_s \alpha_g^2 \left \{1
+\lambda_3(\alpha_u-\alpha_s)+ \omega_3 \frac{1}{2} ( 7 \alpha_g
- 3) \right\} \, , \nonumber\\
V_{\parallel}(\alpha_i,\mu) &=& 120\alpha_u \alpha_s \alpha_g \left(
v_{00}+v_{10}(3\alpha_g-1)\right)\, ,
\nonumber \\
A_{\parallel}(\alpha_i,\mu) &=& 120 \alpha_u \alpha_s \alpha_g
a_{10} (\alpha_s-\alpha_u)\, ,
\nonumber\\
V_{\perp}(\alpha_i,\mu) &=& -30\alpha_g^2
\left\{h_{00}(1-\alpha_g)+h_{01}\left[\alpha_g(1-\alpha_g)-6\alpha_u
\alpha_s\right] \right.  \nonumber\\
&&\left. +h_{10}\left[
\alpha_g(1-\alpha_g)-\frac{3}{2}\left(\alpha_u^2+\alpha_s^2\right)\right]\right\}\,
, \nonumber\\
A_{\perp}(\alpha_i,\mu) &=&  30 \alpha_g^2 (\alpha_u-\alpha_s) \left\{h_{00}+h_{01}\alpha_g+\frac{1}{2}h_{10}(5\alpha_g-3)  \right\}, \nonumber\\
A(u,\mu)&=&6u(1-u)\left\{
\frac{16}{15}+\frac{24}{35}a_2+20\eta_3+\frac{20}{9}\eta_4 \right.
\nonumber \\
&&+\left[
-\frac{1}{15}+\frac{1}{16}-\frac{7}{27}\eta_3\omega_3-\frac{10}{27}\eta_4\right]C^{\frac{3}{2}}_2(2u-1)
\nonumber\\
&&\left.+\left[
-\frac{11}{210}a_2-\frac{4}{135}\eta_3\omega_3\right]C^{\frac{3}{2}}_4(2u-1)\right\}+\left\{
 -\frac{18}{5}a_2+21\eta_4\omega_4\right\} \nonumber\\
 && \left\{2u^3(10-15u+6u^2) \log u+2\bar{u}^3(10-15\bar{u}+6\bar{u}^2) \log \bar{u}
 \right. \nonumber\\
 &&\left. +u\bar{u}(2+13u\bar{u})\right\} \, ,\nonumber\\
 g_K(u,\mu)&=&1+g_2C^{\frac{1}{2}}_2(2u-1)+g_4C^{\frac{1}{2}}_4(2u-1)\, ,\nonumber\\
 B(u,\mu)&=&g_K(u,\mu)-\phi_K(u,\mu)\, ,
\end{eqnarray}
where
\begin{eqnarray}
h_{00}&=&v_{00}=-\frac{\eta_4}{3} \, ,\nonumber\\
a_{10}&=&\frac{21}{8}\eta_4 \omega_4-\frac{9}{20}a_2 \, ,\nonumber\\
v_{10}&=&\frac{21}{8}\eta_4 \omega_4 \, ,\nonumber\\
h_{01}&=&\frac{7}{4}\eta_4\omega_4-\frac{3}{20}a_2 \, ,\nonumber\\
h_{10}&=&\frac{7}{2}\eta_4\omega_4+\frac{3}{20}a_2 \, ,\nonumber\\
g_2&=&1+\frac{18}{7}a_2+60\eta_3+\frac{20}{3}\eta_4 \, ,\nonumber\\
g_4&=&-\frac{9}{28}a_2-6\eta_3\omega_3 \, ,
\end{eqnarray}
 here  $ C_2^{\frac{1}{2}}$, $ C_4^{\frac{1}{2}}$
 and $ C_2^{\frac{3}{2}}$ are Gegenbauer polynomials,
  $\eta_3=\frac{f_{3K}}{f_K}\frac{m_q+m_s}{M_K^2}$ and  $\rho^2={m_s^2\over M_K^2}$
 \cite{LCSR,LCSRreview,Belyaev94,Ball98,Ball06}.

The explicit expressions of the Borel transformed correlation
functions $B_M\Pi^1_\mu$ and $B_M\Pi^2_\mu$ in the hadronic
representation,
\begin{eqnarray}
B_M\Pi^1_\mu
  &=&\left\{\frac{ G_{D^* D_sK} m_{D^*}f_{D^*}f_{D_s}m_{D_s}^2}
  {m_c+m_s } \frac{m_{D^*}^2-m_{D_s}^2+m_K^2}{m_{D^*}^2}\exp\left\{-\frac{m_{D^*}^2}{M_1^2}-\frac{m_{D_s}^2}{M_2^2}\right\}\right. \nonumber\\
  &&\left.+  \frac{ G_{D_0D_s K} f_{D_0}f_{D_s}m_{D_s}^2  }
  {m_c+m_s}\exp\left\{-\frac{m_{D_0}^2}{M_1^2}-\frac{m_{D_s}^2}{M_2^2}\right\}\right\}q_\mu+ \nonumber \\
&&+ \left\{-\frac{ G_{D^* D_sK} m_{D^*}f_{D^*}f_{D_s}m_{D_s}^2}
  {m_c+m_s }\frac{m_{D^*}^2+m_{D_s}^2-m_K^2}{m_{D^*}^2}\exp\left\{-\frac{m_{D^*}^2}{M_1^2}-\frac{m_{D_s}^2}{M_2^2}\right\}\right. \nonumber\\
  &&\left.+  \frac{ G_{D_0D_s K} f_{D_0}f_{D_s}m_{D_s}^2  }
  { m_c+m_s }\exp\left\{-\frac{m_{D_0}^2}{M_1^2}-\frac{m_{D_s}^2}{M_2^2}\right\}\right\}P_\mu+ \cdots  ,
\end{eqnarray}
\begin{eqnarray}
B_M\Pi^2_\mu &=&\left\{\frac{ G_{D^*_{s}DK} m_{D^*_s}f_{D^*_s}f_D
m_D^2}
  {m_c+m_u}\frac{m_{D^*_s}^2-m_D^2+m_K^2}{m_{D^*_s}^2}\exp\left\{-\frac{m_{D_s^*}^2}{M_1^2}-\frac{m_D^2}{M_2^2}\right\}\right. \nonumber\\
  &&+\left.  \frac{ G_{D_{s0}DK} f_{D_{s0}}f_{D}m_D^2  }
  {m_c+m_u}\exp\left\{-\frac{m_{D_{s0}}^2}{M_1^2}-\frac{m_D^2}{M_2^2}\right\} \right\}q_\mu +\nonumber\\
&&\left\{-\frac{ G_{D^*_{s}DK} m_{D^*_s}f_{D^*_s}f_D m_D^2}
  {m_c+m_u}\frac{m_{D^*_s}^2+m_D^2-m_K^2}{m_{D^*_s}^2}\exp\left\{-\frac{m_{D_s^*}^2}{M_1^2}-\frac{m_D^2}{M_2^2}\right\}\right. \nonumber\\
  &&+\left. \frac{ G_{D_{s0}DK} f_{D_{s0}}f_{D}m_D^2  }
  {m_c+m_u}\exp\left\{-\frac{m_{D_{s0}}^2}{M_1^2}-\frac{m_D^2}{M_2^2}\right\} \right\}P_\mu + \cdots  ,
\end{eqnarray}
here we have not shown  the contributions from the high resonances
and continuum states  explicitly for simplicity.

The explicit expressions of the Borel transformed correlation
functions $B_M\Pi^1_\mu$ and $B_M\Pi^2_\mu$ at the level of
quark-gluon degrees of freedom,
\begin{eqnarray}
B_{M}\Pi_\mu^1&=&q_\mu
\exp\left\{-\frac{u_0(1-u_0)m_K^2+m_c^2}{M^2}\right\}\left\{
\frac{f_Km_K^2M^2}{m_s}\left[\phi_p(u_0)-\frac{d}{6du_0}\phi_\sigma(u_0)\right]\right. \nonumber\\
&&+m_cf_Km_K^2\int_0^{u_0}dtB(t)-f_{3K}m_K^2 \int_0^{u_0}d\alpha_s
\int_{u_0-\alpha_s}^{1-\alpha_s}d\alpha_g
\frac{\left[2(\alpha_s-u_0)-\alpha_g\right]T(\alpha_u,\alpha_g,\alpha_s)}{\alpha_g^2}
\nonumber\\
&&-\frac{2m_cf_Km_K^4}{M^2} \int_{1-u_0}^1
d\alpha_g\int_0^{\alpha_g} d\beta\int_0^{1-\beta}d\alpha
\frac{(1-u_0)\Phi(1-\alpha-\beta,\beta,\alpha)}{\alpha_g^2}
\nonumber \\
&&+\frac{2m_cf_Km_K^4}{M^2} \left[\int_0^{1-u_0}
d\alpha_g\int^{u_0}_{u_0-\alpha_g} d\alpha_s
 \int_0^{\alpha_s}d\alpha +\int_{1-u_0}^1
d\alpha_g\int_{u_0-\alpha_g}^{1-\alpha_g} d\alpha_s
 \int_{0}^{\alpha_s}d\alpha\right] \nonumber\\
 &&\left.\frac{\Phi(1-\alpha-\alpha_g,\alpha_g,\alpha)}{\alpha_g}\right\}
\nonumber \\
&+&P_\mu
\exp\left\{-\frac{u_0(1-u_0)m_K^2+m_c^2}{M^2}\right\}\left\{\frac{f_Km_K^2M^2}{m_s}
u_0\phi_p(u_0) +m_cf_Km_K^2 u_0\int_0^{u_0}dt B(t) \right. \nonumber \\
&&-\frac{f_Km_K^2}{6m_s}\left[
\phi_\sigma(u_0)M^2+\frac{d}{du_0}u_0\phi_\sigma(u_0)M^2+2m_c^2\phi_\sigma(u_0)
\right] \nonumber\\
&&+m_cf_K  \left[ -\phi_K(u_0)M^2 +\frac{m_K^2m_c^2A(u_0)}{4M^2}
\right] \nonumber\\
&&+3u_0f_{3K}m_K^2\int_0^{u_0}d\alpha_s
\int_{u_0-\alpha_s}^{1-\alpha_s}d\alpha_g
\frac{T(\alpha_u,\alpha_g,\alpha_s)}{\alpha_g} \nonumber\\
&&+2f_{3K}\left[M^2\frac{d}{du_0}-u_0m_K^2\right]\int_0^{u_0}d\alpha_s
\int_{u_0-\alpha_s}^{1-\alpha_s}d\alpha_g
\frac{\alpha_g+\alpha_s-u_0}{\alpha_g^2}T(\alpha_u,\alpha_g,\alpha_s) \nonumber\\
 &&-\frac{2u_0m_cf_Km_K^4}{M^2} \int_{1-u_0}^1 d\alpha_g\int_0^{\alpha_g}
d\beta\int_0^{1-\beta}d\alpha
\frac{(1-u_0)\Phi(1-\alpha-\beta,\beta,\alpha)}{\alpha_g^2}
\nonumber \\
&&+\frac{2u_0m_cf_Km_K^4}{M^2}\left[\int_0^{1-u_0}
d\alpha_g\int^{u_0}_{u_0-\alpha_g} d\alpha_s
 \int_0^{\alpha_s}d\alpha +\int_{1-u_0}^1
d\alpha_g\int_{u_0-\alpha_g}^{1-\alpha_g} d\alpha_s
 \int_{0}^{\alpha_s}d\alpha\right] \nonumber\\
 && \left. \frac{\Phi(1-\alpha-\alpha_g,\alpha_g,\alpha)}{\alpha_g} +m_cf_Km_K^2 \int_0^{u_0}
d\alpha_s\int_{u_0-\alpha_s}^{1-\alpha_s} d\alpha_g
\frac{\Psi(\alpha_u,\alpha_g,\alpha_s)} {\alpha_g}\right\}\, , \\
B_M \Pi_\mu^2&=&B_M \Pi_\mu^1(u\longleftrightarrow 1-u; \alpha_s
\longleftrightarrow \alpha_u) \, ,
\end{eqnarray}
here $u_0=\frac{M_1^2}{M_1^2+M_2^2}$, $M^2=\frac{M_1^2
M_2^2}{M_1^2+M_2^2}$.

  The explicit expressions of the notations $AA$, $BB$, $CC$,
$DD$, $EE$ and $FF$,
\begin{eqnarray}
AA&=&\left\{\exp\left\{-\frac{u_0(1-u_0)m_K^2+m_c^2}{M^2}\right\}-\exp\left\{-\frac{s^0_{D^*}}{M^2}\right\}\right\}\left\{\frac{f_Km_K^2M^2}{m_s}
u_0\phi_p(u_0) \right. \nonumber \\
&&\left.- m_cf_K \phi_K(u_0)M^2-\frac{f_Km_K^2}{6m_s}\left[
\phi_\sigma(u_0)M^2+\frac{d}{du_0}u_0\phi_\sigma(u_0)M^2+2m_c^2\phi_\sigma(u_0)
\right] \right\}\nonumber\\
&&+\exp\left\{-\frac{u_0(1-u_0)m_K^2+m_c^2}{M^2}\right\}\left\{m_cf_Km_K^2
u_0\int_0^{u_0}dt B(t) +\frac{f_Km_K^2m_c^3A(u_0)}{4M^2}
 \right. \nonumber \\
&&+3u_0f_{3K}m_K^2\int_0^{u_0}d\alpha_s
\int_{u_0-\alpha_s}^{1-\alpha_s}d\alpha_g
\frac{T(\alpha_u,\alpha_g,\alpha_s)}{\alpha_g} \nonumber\\
&&+2f_{3K}\left[M^2\frac{d}{du_0}-u_0m_K^2\right]\int_0^{u_0}d\alpha_s
\int_{u_0-\alpha_s}^{1-\alpha_s}d\alpha_g
\frac{\alpha_g+\alpha_s-u_0}{\alpha_g^2}T(\alpha_u,\alpha_g,\alpha_s) \nonumber\\
 &&-\frac{2u_0m_cf_Km_K^4}{M^2} \int_{1-u_0}^1 d\alpha_g\int_0^{\alpha_g}
d\beta\int_0^{1-\beta}d\alpha
\frac{(1-u_0)\Phi(1-\alpha-\beta,\beta,\alpha)}{\alpha_g^2}
\nonumber \\
&&+\frac{2u_0m_cf_Km_K^4}{M^2}\left[\int_0^{1-u_0}
d\alpha_g\int^{u_0}_{u_0-\alpha_g} d\alpha_s
 \int_0^{\alpha_s}d\alpha +\int_{1-u_0}^1
d\alpha_g\int_{u_0-\alpha_g}^{1-\alpha_g} d\alpha_s
 \int_{0}^{\alpha_s}d\alpha\right] \nonumber\\
 && \left. \frac{\Phi(1-\alpha-\alpha_g,\alpha_g,\alpha)}{\alpha_g} +m_cf_Km_K^2 \int_0^{u_0}
d\alpha_s\int_{u_0-\alpha_s}^{1-\alpha_s} d\alpha_g
\frac{\Psi(\alpha_u,\alpha_g,\alpha_s)} {\alpha_g}\right\} \, ;
\end{eqnarray}

\begin{eqnarray}
BB&=&\left\{\exp\left\{-\frac{u_0(1-u_0)m_K^2+m_c^2}{M^2}\right\}-\exp\left\{-\frac{s^0_{D^*_s}}{M^2}\right\}\right\}\left\{\frac{f_Km_K^2M^2}{m_s}
u_0\phi_p(u_0)\right. \nonumber \\
&&\left.- m_cf_K \phi_K(u_0)M^2 -\frac{f_Km_K^2}{6m_s}\left[
\phi_\sigma(u_0)M^2+\frac{d}{du_0}u_0\phi_\sigma(u_0)M^2+2m_c^2\phi_\sigma(u_0)
\right] \right\}\nonumber\\
&&+\exp\left\{-\frac{u_0(1-u_0)m_K^2+m_c^2}{M^2}\right\}\left\{m_cf_Km_K^2
u_0\int_0^{u_0}dt B(t) +\frac{f_Km_K^2m_c^3A(u_0)}{4M^2}
 \right. \nonumber \\
&&+3u_0f_{3K}m_K^2\int_0^{u_0}d\alpha_s
\int_{u_0-\alpha_s}^{1-\alpha_s}d\alpha_g
\frac{T(\alpha_u,\alpha_g,\alpha_s)}{\alpha_g} \nonumber\\
&&+2f_{3K}\left[M^2\frac{d}{du_0}-u_0m_K^2\right]\int_0^{u_0}d\alpha_s
\int_{u_0-\alpha_s}^{1-\alpha_s}d\alpha_g
\frac{\alpha_g+\alpha_s-u_0}{\alpha_g^2}T(\alpha_u,\alpha_g,\alpha_s) \nonumber\\
 &&-\frac{2u_0m_cf_Km_K^4}{M^2} \int_{1-u_0}^1 d\alpha_g\int_0^{\alpha_g}
d\beta\int_0^{1-\beta}d\alpha
\frac{(1-u_0)\Phi(1-\alpha-\beta,\beta,\alpha)}{\alpha_g^2}
\nonumber \\
&&+\frac{2u_0m_cf_Km_K^4}{M^2}\left[\int_0^{1-u_0}
d\alpha_g\int^{u_0}_{u_0-\alpha_g} d\alpha_s
 \int_0^{\alpha_s}d\alpha +\int_{1-u_0}^1
d\alpha_g\int_{u_0-\alpha_g}^{1-\alpha_g} d\alpha_s
 \int_{0}^{\alpha_s}d\alpha\right] \nonumber\\
 && \left. \frac{\Phi(1-\alpha-\alpha_g,\alpha_g,\alpha)}{\alpha_g} +m_cf_Km_K^2 \int_0^{u_0}
d\alpha_s\int_{u_0-\alpha_s}^{1-\alpha_s} d\alpha_g
\frac{\Psi(\alpha_u,\alpha_g,\alpha_s)}
{\alpha_g}\right\}\mid_{\alpha_s \leftrightarrow \alpha_u} \, ;
\nonumber \\
\end{eqnarray}

\begin{eqnarray}
CC&=&
\left\{\exp\left\{-\frac{u_0(1-u_0)m_K^2+m_c^2}{M^2}\right\}-\exp\left\{-\frac{s^0_{D_0}}{M^2}\right\}\right\}
\frac{f_Km_K^2M^2}{m_s}\left\{\phi_p(u_0)-\frac{d}{6du_0}\phi_\sigma(u_0)\right\} \nonumber\\
&&+ \exp\left\{-\frac{u_0(1-u_0)m_K^2+m_c^2}{M^2}\right\}\left\{
 m_cf_Km_K^2\int_0^{u_0}dtB(t) \right. \nonumber\\
&&-f_{3K}m_K^2 \int_0^{u_0}d\alpha_s
\int_{u_0-\alpha_s}^{1-\alpha_s}d\alpha_g
\frac{\left[2(\alpha_s-u_0)-\alpha_g\right]T(\alpha_u,\alpha_g,\alpha_s)}{\alpha_g^2}
\nonumber\\
&&-\frac{2m_cf_Km_K^4}{M^2} \int_{1-u_0}^1
d\alpha_g\int_0^{\alpha_g} d\beta\int_0^{1-\beta}d\alpha
\frac{(1-u_0)\Phi(1-\alpha-\beta,\beta,\alpha)}{\alpha_g^2}
\nonumber \\
&&+\frac{2m_cf_Km_K^4}{M^2} \left[\int_0^{1-u_0}
d\alpha_g\int^{u_0}_{u_0-\alpha_g} d\alpha_s
 \int_0^{\alpha_s}d\alpha +\int_{1-u_0}^1
d\alpha_g\int_{u_0-\alpha_g}^{1-\alpha_g} d\alpha_s
 \int_{0}^{\alpha_s}d\alpha\right] \nonumber\\
 &&\left.\frac{\Phi(1-\alpha-\alpha_g,\alpha_g,\alpha)}{\alpha_g}\right\}
\, ;
\end{eqnarray}

\begin{eqnarray}
DD&=&
\left\{\exp\left\{-\frac{u_0(1-u_0)m_K^2+m_c^2}{M^2}\right\}-\exp\left\{-\frac{s^0_{D_{s0}}}{M^2}\right\}\right\}
\frac{f_Km_K^2M^2}{m_s}\left\{\phi_p(u_0)-\frac{d}{6du_0}\phi_\sigma(u_0)\right\} \nonumber\\
&&+ \exp\left\{-\frac{u_0(1-u_0)m_K^2+m_c^2}{M^2}\right\}\left\{
 m_cf_Km_K^2\int_0^{u_0}dtB(t) \right. \nonumber\\
&&-f_{3K}m_K^2 \int_0^{u_0}d\alpha_s
\int_{u_0-\alpha_s}^{1-\alpha_s}d\alpha_g
\frac{\left[2(\alpha_s-u_0)-\alpha_g\right]T(\alpha_u,\alpha_g,\alpha_s)}{\alpha_g^2}
\nonumber\\
&&-\frac{2m_cf_Km_K^4}{M^2} \int_{1-u_0}^1
d\alpha_g\int_0^{\alpha_g} d\beta\int_0^{1-\beta}d\alpha
\frac{(1-u_0)\Phi(1-\alpha-\beta,\beta,\alpha)}{\alpha_g^2}
\nonumber \\
&&+\frac{2m_cf_Km_K^4}{M^2} \left[\int_0^{1-u_0}
d\alpha_g\int^{u_0}_{u_0-\alpha_g} d\alpha_s
 \int_0^{\alpha_s}d\alpha +\int_{1-u_0}^1
d\alpha_g\int_{u_0-\alpha_g}^{1-\alpha_g} d\alpha_s
 \int_{0}^{\alpha_s}d\alpha\right] \nonumber\\
 &&\left.\frac{\Phi(1-\alpha-\alpha_g,\alpha_g,\alpha)}{\alpha_g}\right\} \mid_{\alpha_s \leftrightarrow \alpha_u}
\, ;
\end{eqnarray}

\begin{eqnarray}
EE&=&\left\{\exp\left\{-\frac{u_0(1-u_0)m_K^2+m_c^2}{M^2}\right\}-\exp\left\{-\frac{s^0_{D_0}}{M^2}\right\}\right\}\left\{\frac{f_Km_K^2M^2}{m_s}
u_0\phi_p(u_0) \right. \nonumber \\
&&\left.- m_cf_K \phi_K(u_0)M^2-\frac{f_Km_K^2}{6m_s}\left[
\phi_\sigma(u_0)M^2+\frac{d}{du_0}u_0\phi_\sigma(u_0)M^2+2m_c^2\phi_\sigma(u_0)
\right] \right\}\nonumber\\
&&+\exp\left\{-\frac{u_0(1-u_0)m_K^2+m_c^2}{M^2}\right\}\left\{m_cf_Km_K^2
u_0\int_0^{u_0}dt B(t) +\frac{f_Km_K^2m_c^3A(u_0)}{4M^2}
 \right. \nonumber \\
&&+3u_0f_{3K}m_K^2\int_0^{u_0}d\alpha_s
\int_{u_0-\alpha_s}^{1-\alpha_s}d\alpha_g
\frac{T(\alpha_u,\alpha_g,\alpha_s)}{\alpha_g} \nonumber\\
&&+2f_{3K}\left[M^2\frac{d}{du_0}-u_0m_K^2\right]\int_0^{u_0}d\alpha_s
\int_{u_0-\alpha_s}^{1-\alpha_s}d\alpha_g
\frac{\alpha_g+\alpha_s-u_0}{\alpha_g^2}T(\alpha_u,\alpha_g,\alpha_s) \nonumber\\
 &&-\frac{2u_0m_cf_Km_K^4}{M^2} \int_{1-u_0}^1 d\alpha_g\int_0^{\alpha_g}
d\beta\int_0^{1-\beta}d\alpha
\frac{(1-u_0)\Phi(1-\alpha-\beta,\beta,\alpha)}{\alpha_g^2}
\nonumber \\
&&+\frac{2u_0m_cf_Km_K^4}{M^2}\left[\int_0^{1-u_0}
d\alpha_g\int^{u_0}_{u_0-\alpha_g} d\alpha_s
 \int_0^{\alpha_s}d\alpha +\int_{1-u_0}^1
d\alpha_g\int_{u_0-\alpha_g}^{1-\alpha_g} d\alpha_s
 \int_{0}^{\alpha_s}d\alpha\right] \nonumber\\
 && \left. \frac{\Phi(1-\alpha-\alpha_g,\alpha_g,\alpha)}{\alpha_g} +m_cf_Km_K^2 \int_0^{u_0}
d\alpha_s\int_{u_0-\alpha_s}^{1-\alpha_s} d\alpha_g
\frac{\Psi(\alpha_u,\alpha_g,\alpha_s)} {\alpha_g}\right\} \, ;
\end{eqnarray}

\begin{eqnarray}
FF&=&\left\{\exp\left\{-\frac{u_0(1-u_0)m_K^2+m_c^2}{M^2}\right\}-\exp\left\{-\frac{s^0_{D_{s0}}}{M^2}\right\}\right\}\left\{\frac{f_Km_K^2M^2}{m_s}
u_0\phi_p(u_0) \right. \nonumber \\
&&\left.- m_cf_K \phi_K(u_0)M^2-\frac{f_Km_K^2}{6m_s}\left[
\phi_\sigma(u_0)M^2+\frac{d}{du_0}u_0\phi_\sigma(u_0)M^2+2m_c^2\phi_\sigma(u_0)
\right] \right\}\nonumber\\
&&+\exp\left\{-\frac{u_0(1-u_0)m_K^2+m_c^2}{M^2}\right\}\left\{m_cf_Km_K^2
u_0\int_0^{u_0}dt B(t) +\frac{f_Km_K^2m_c^3A(u_0)}{4M^2}
 \right. \nonumber \\
&&+3u_0f_{3K}m_K^2\int_0^{u_0}d\alpha_s
\int_{u_0-\alpha_s}^{1-\alpha_s}d\alpha_g
\frac{T(\alpha_u,\alpha_g,\alpha_s)}{\alpha_g} \nonumber\\
&&+2f_{3K}\left[M^2\frac{d}{du_0}-u_0m_K^2\right]\int_0^{u_0}d\alpha_s
\int_{u_0-\alpha_s}^{1-\alpha_s}d\alpha_g
\frac{\alpha_g+\alpha_s-u_0}{\alpha_g^2}T(\alpha_u,\alpha_g,\alpha_s) \nonumber\\
 &&-\frac{2u_0m_cf_Km_K^4}{M^2} \int_{1-u_0}^1 d\alpha_g\int_0^{\alpha_g}
d\beta\int_0^{1-\beta}d\alpha
\frac{(1-u_0)\Phi(1-\alpha-\beta,\beta,\alpha)}{\alpha_g^2}
\nonumber \\
&&+\frac{2u_0m_cf_Km_K^4}{M^2}\left[\int_0^{1-u_0}
d\alpha_g\int^{u_0}_{u_0-\alpha_g} d\alpha_s
 \int_0^{\alpha_s}d\alpha +\int_{1-u_0}^1
d\alpha_g\int_{u_0-\alpha_g}^{1-\alpha_g} d\alpha_s
 \int_{0}^{\alpha_s}d\alpha\right] \nonumber\\
 && \left. \frac{\Phi(1-\alpha-\alpha_g,\alpha_g,\alpha)}{\alpha_g} +m_cf_Km_K^2 \int_0^{u_0}
d\alpha_s\int_{u_0-\alpha_s}^{1-\alpha_s} d\alpha_g
\frac{\Psi(\alpha_u,\alpha_g,\alpha_s)} {\alpha_g}\right\}
\mid_{\alpha_s \leftrightarrow \alpha_u}\, . \nonumber \\
\end{eqnarray}
\section*{Acknowledgments}
This  work is supported by National Natural Science Foundation,
Grant Number 10405009,  and Key Program Foundation of NCEPU.

\end{document}